# Do Black and Indigenous Communities Receive their Fair Share of Vaccines Under the 2018 CDC Guidelines?[1]

Parag A. Pathak[2], Harald Schmidt[3], Adam Solomon[4],
Edwin Song[5], Tayfun Sönmez[6], M. Utku Ünver[7]

September 5, 2020

**Abstract**. A major focus of debate about rationing guidelines for COVID-19 vaccines is whether and how to prioritize access for minority populations that have been particularly affected by the pandemic, and been the subject of historical and structural disadvantage, particularly Black and Indigenous individuals. We simulate the 2018 CDC Vaccine Allocation guidelines using data from the American Community Survey under different assumptions on total vaccine supply. Black and Indigenous individuals combined receive a higher share of vaccines compared to their population share for all assumptions on total vaccine supply. However, their vaccine share under the 2018 CDC guidelines is considerably lower than their share of COVID-19 deaths and age-adjusted deaths. We then simulate one method to incorporate disadvantage in vaccine allocation via a reserve system. In a reserve system, units are placed into categories and units reserved for a category give preferential treatment to individuals from that category. Using the Area Deprivation Index (ADI) as a proxy for disadvantage, we show that a 40% high-ADI reserve increases the number of vaccines allocated to Black or Indigenous individuals, with a share that approaches their COVID-19 death share when there are about 75 million units. Our findings illustrate that whether an allocation is equitable depends crucially on the benchmark and highlight the importance of considering the expected distribution of outcomes from implementing vaccine allocation guidelines.

[1] We are grateful to Chetan Patel for excellent assistance.
[2] MIT, Cambridge MA 02139, USA. ppathak@mit.edu
[3] University of Pennsylvania, Philadelphia, PA 19104, USA. schmidth@upenn.edu
[4] MIT, Cambridge MA 02139, USA. adamsol@mit.edu
[5] MIT, Cambridge MA 02139, USA. song22@mit.edu
[6] Boston College, Chestnut Hill MA 02467, USA. tayfun.sonmez@bc.edu
[7] Boston College, Chestnut Hill MA 02467, USA. unver@bc.edu.

## I. Introduction

There is widespread evidence that the burden in terms of unemployment, hospitalizations, and deaths due to COVID-19 is disproportionately higher in minority populations, particularly Black communities, as well as American Indian or Alaska Natives (hereafter, "Indigenous") (see, e.g., Tai, et al. 2020). Current estimates are that Blacks are dying at 2.4 times the rate of whites and Indigenous are dying at 1.5 times the rate of Whites (Covidtracking 2020).

This reality has led many to argue that the hardest-hit racial and ethnic groups should receive priority access to COVID-19 vaccines. For example, Melinda Gates, a prominent funder of vaccine development efforts, has suggested that after health care workers, "Black people [should be] next, and many other people of color" (Ducharme 2020). Likewise, Schmidt (2020) articulates reasons why rationing approaches should give priority to groups that have been structurally and historically disadvantaged.

Others have argued that a rationing policy that prioritizes disadvantaged groups would be inefficient and unfair. In public comments to the National Academy of Sciences, Engineering, and Medicine (NASEM) Committee on Vaccine Allocation, Dr. Marc Lipsitch raised the possibility that if herd immunity develops in disadvantaged communities due to higher incidence, then these patients may be immune and therefore would not need prioritized vaccines (Lipsitch 2020). There is also the concern that without public trust in a vaccine, disadvantaged communities may believe they are being used as 'guinea pigs' if given priority (Toner et al. 2020). Some question whether allocation based on race or ethnicity would be legal or equitable and undermine broader trust in the health system (see, e.g., McCaughey 2020, Twohey 2020).

Furthermore, the ACIP COVID-19 Working Group does not consider race or ethnicity in their August 26, 2020 interim framework for COVID-19 vaccine allocation (Dooling 2020). A draft report on equitable vaccine allocation by NASEM published on September 2, 2020 emphasized that the allocation framework had to be sensitive to the disproportionate impact on minority populations. Instead of directly using race or ethnicity as a criterion for allocation, the report proposed to allocate within each of the subgroups of populations in four priority tiers by their level of deprivation, as captured in the CDC's Social Vulnerability Index (however, race and ethnicity are included in the 15 criteria in this index). Guidance issued to states by the CDC in the week prior to the publication of the NASEM report recommends drawing on ACIP and NASEM's eventual conclusions, but also encourages that in identifying priority populations, planners should consider "people from racial and ethnic minority populations" as one of several subgroups at increased risk for severe illness (Kaplan, Wu, and Thomas 2020).

The issue of prioritizing racial and ethnic minority populations raises two important questions. First, under existing guidelines, do minority communities receive a fair share of vaccines (however "fair" is operationalized)? Second, how would one implement a priority for racial and ethnic minority populations in practice?

This paper examines the allocation of vaccines to disadvantaged communities and across US states under the 2018 CDC Influenza Vaccine Allocation guidelines. We focus on these guidelines since, at the time of writing, all proposed allocation frameworks by ACIP, NASEM, and the CDC are in considerable flux. The 2018 CDC Vaccine guidelines underwent comprehensive consultation and review, and as such are an important precedent and benchmark for on-going discussions about allocation of a COVID-19 Vaccine (Cohen 2020, Splete 2020). While a comprehensive analysis of the impact on minorities would, of course, require considering all types, including ethnic subtypes, given the complexity of the issue at hand and our overall goals, we focus the discussion here on allocations to Black (both Hispanic and non-Hispanic) and Indigenous populations relative to white ones (both Hispanic and non-Hispanic), by reference to the respective population size, COVID-19 related deaths, and age-adjusted deaths using data from the American Community Survey. For ease of illustration, throughout, we combine data for Black and Indigenous groups. We then consider a suggestion by Schmidt (2020) to use the Area Deprivation Index (ADI), an index made available by Kind and Buckingham (2018), which ranks neighborhoods by socioeconomic status disadvantage, as part of the allocation system. To implement priority for hardest-hit groups, we use a reserve system proposed by Pathak, Sönmez, Ünver, and Yenmez (2020a,b). For pragmatic reasons, we simulate by how much a 20% and 40% high-ADI reserve with a gradient cut-off as described below, would change vaccine access for Black and Indigenous individuals and how an ADI reserve would influence distribution across states.

## II. The 2018 CDC Influenza Vaccine Guidelines

In 2018, CDC guidelines place patients into five tiers based on occupation and demographic factors indicating risk for contracting or spreading influenza. Table 1, from the CDC guidelines, contains the definitions of the tiers. Tier 1 includes essential occupational groups and high-risk populations such as pregnant women and infants and toddlers. Individuals are placed in Tiers 2-4 based on the occupational groups and risk status, shown in the table. Tier 5 is the largest group, consisting of healthy adults between 19-64 years old. If there are not enough vaccines for those in the first tier, the first tier is further portioned into seven ranked sub-tiers. For each of Tiers 2-5, all those in the same tier have equal priority.

To estimate the number of patients in each tier, we need information on the joint distribution of all individual characteristics used to identify the tier in which an individual belongs. As far as know, such data do not exist. We therefore use the American Community Survey (ACS) 2014-2018 5-year Public Use Microdata Sample (PUMS) to identify patients eligible for each tier. We combine this with information from the 2018 Behavioral Risk Factor Surveillance System (BRFSS) data to identify information on high-risk conditions.

For many of the CDC's occupation-based groups, the PUMS data does not have adequate detail to definitively identify an individual as a member of the group. For example, one of the CDC's groups consists of manufacturing staff of pandemic vaccine and antiviral drugs. The PUMS data identifies those working in the pharmaceutical manufacturing industry, but provides no additional detail. Our approach in such a situation is to construct a superset of the group (such as all those working in pharmaceutical manufacturing) and then randomly and independently assign each observation in the superset to the group with a fixed probability for the group. The fixed probability is calibrated so that the number of people in the US population represented by the observations assigned to the group is approximately equal to the number of people in the group as confirmed or estimated from a secondary source.

The PUMS data does not contain information on the health characteristics of surveyed individuals. We define an individual as high risk if they have a comorbidity such as cancer, kidney disease, heart disease, diabetes, or high BMI in the BFRSS. We then group the BRFSS data by age bin, sex, race, and for each combination, we compute the proportion of the population that is high-risk. Then we return to the PUMS data, and each individual is assigned to the "high risk" category with the probability relevant to their group in the BRFSS.

Finally, we must impute an ADI score for each individual. We cannot assign the actual ADI of each person in the PUMS data because ADI is computed using census block level averages and the PUMS geographic specificity is only at the Public Use Microdata Area (PUMA) level. Singh (2003) reports the 17 components and their coefficients used to determine the census tract level ADI in 1990. We use these coefficients to find ADI at the household level. Each coefficient is assigned a weight for the household using PUMS data at the household level. The weighted sum of coefficients gives the raw numerical ADI score of the household. Since PUMS links households to individuals, we assign these scores to each individual of the household in PUMS. We divide the distribution in deciles and assign each decile scores 1 to 10.

We are mindful that there are several limitations to our simulation because of data constraints. The most important limitation stem from certain data omissions in PUMS data and the fact that we compute ADI at the family level rather than the census block level. We omit 8 million people living in group quarters such as prisons, nursing facilities, and college and university housing as most PUMS variables do not exist for them and they are automatically excluded from the high-ADI group.[8] Some missing variables for people living in household setting are assumed to be 0. The Appendix provides more detail on our methodology.

## III. Vaccine Allocation under 2018 CDC guidelines

Figure 1 reports the proportion of vaccines allocated to Black and Indigenous populations (combined) as we vary the total doses of vaccines under the 2018 CDC guidelines assuming a single type of vaccine. The five tiers are demarcated as the number of vaccines increases. We estimate that there are just under 25 million individuals in Tier 1.[9] Figure 1 shows that after allocating these 25 million units, about 18% of vaccines will be allocated to Black and Indigenous recipients in Tier 2. The proportion of vaccines allocated plateaus at 18%, until it converges to the population fraction Black and Indigenous at which point there are enough vaccines to accommodate the entire US population of about 325 million.

[8] ADI computed at the census block level also omits computing an ADI for census blocks with more than 33% of the population living in group quarters.

[9] The CDC guidelines estimate about 26 million individuals are in Tier 1. We find 25 million in Tier 1 because of a difference in the count of pregnant women. The CDC guidelines estimate 4 million pregnant women in the United States. Our estimate using the ACS is the number of people who gave birth in the last year multiplied by the length of a pregnancy as a fraction of a year results in 3 million.

We report three benchmarks for comparison in the Figure: the population fraction of Black and Indigenous groups, the fraction of COVID-related deaths among Black and Indigenous populations, and the age-adjusted fraction of COVID deaths among them. Data on death rates and age-adjusted death rates come from APM Research Lab's Color of Coronavirus database, and are as of August 5, 2020. Figure 1 shows that while 15.5% of the US population is Black or Indigenous, the two groups account for 24.6% of COVID deaths and 28.7% of COVID age-adjusted death.

The proportion of vaccines allocated to Black and Indigenous recipients is above their population share, but well below their share of COVID deaths or age-adjusted deaths. Therefore, when judged compared to their population share, Black and Indigenous individuals receive a larger share of doses for all scenarios on vaccine supply. However, this comparison by itself does not settle the question of whether the allocation is fair, and rather, draws attention to the justification of the reference point. When compared to the share of COVID deaths or age-adjusted deaths, the proportion of vaccines allocated to Black and Indigenous recipients is well below these benchmarks. If there are 100 million units of a vaccine, under the CDC guidelines, about 17.5% units would go to Black and Indigenous individuals, while they represent 28.7% of COVID age-adjusted deaths. As a result, whether an allocation is judged to be fair depends crucially on the benchmark used for comparisons.

Figure 2 reports on the share of vaccines allocated to each state under the 2018 CDC guidelines for two scenarios on vaccine supply: 20 million and 50 million total vaccines. At 20 million units, the CDC priority system would allocate units partway through Tier 1, while at 50 million units the CDC priority system would allocate units partway through Tier 3. For each assumption on the total number of vaccines, the Figure shows the state share of vaccines compared to three benchmarks: the state population share, the state case share, and the state death share. For each normalization, a value of one means that the state share of vaccines is equal to the population share, case share or death share. A value greater than one means that the state obtains a greater share of vaccines relative to the benchmark, while a value less than one means the state obtains a smaller share relative to the benchmark.

The top panel of Figure 2 shows that relative to population, states like California, Florida, and Michigan, receive a smaller share of vaccines relative to their population. This remains true when the number of vaccines increases to 50 million units. At 50 million units, several northeastern states receive a ratio less than one, while their ratio was greater than one at 20 million units. For example, at 20 million units, Massachusetts's share of vaccines is 1.02, while at 50 million units, its share is 0.94.

The second panel of Figure 2 shows that relative to the state case share, southern states receive much less than their proportionate share, while mountain and midwestern states receive more than their proportionate share. The comparison relative to state population share in the top panel did not indicate major differences between these states. For example, Louisiana's share of vaccines relative to its share of cases is 0.59 at 20 million units and 0.58 at 50 million units. Relative to state population, Louisiana is at 1.06 and 1.04, respectively. Montana's share of vaccines relative to case share is over 3 at both levels of vaccine supply, but it is close to one when judged by population share. For the state case share benchmark, there is relatively little difference across states as vaccine supply increases. This can be seen by comparing the colors across the two scenarios in the second panel of Figure 2.

The third panel of Figure 2 shows that relative to the share of deaths in the state, northeastern states receives less than their share of deaths, while southeastern states receive more than their share of deaths. New Jersey and New York stand out under both the 20 and 50 million scenarios as receiving the least compared to the share of deaths. Large states like California and Texas receive more than their share of vaccines relative to share of deaths.

The variation across these three metrics illustrates that how an equitable allocation is defined matters a great deal for regional comparisons.

## IV. Prioritizing the Hardest-Hit in the 2018 CDC system

### A. How to Prioritize?

There is no explicit consideration for prioritizing hardest-hit groups in the 2018 CDC vaccine allocation system. Pathak, et. al (2020a,b) illustrate that despite widespread consensus on the desirability of incorporating multiple ethical criteria into an allocation system, many medical rationing guidelines struggle to do so because they employ a priority system. If hardest-hit individuals receive heightened priority, there is a potential that all units will go to the group with heightened priority, leaving no units available for the rest of society. For example, this is a major challenge in formulating guidelines of rationing ventilators during influenza pandemics, with some states like New York and Minnesota abandoning priority for essential personnel, and a few states like Michigan adopting essential personnel priority. Several other states are vague about implementation (see Pathak et. al. 2020b for details).

Imprecision on vaccine allocation is a major hurdle for public acceptance of any rationing scheme. The inability to accommodate various desirable ethical goals may be a consequence of the mechanical structure of a priority system that allocates all units using the same priority order of individuals. A more flexible alternative is a reserve system, which divides resources into multiple categories and uses a potentially different criteria for allocation of units in each category. These category-specific criteria reflect the balance between ethical values guiding allocation of units in the category.

If units are placed into a reserve for hardest-hit groups, there are two important remaining questions: how are the hardest-hit defined and what share of units will be placed into the category in which the hardest-hit obtain preferential treatment?

We focus on using the Area Deprivation Index (ADI) to identify hardest-hit areas, following the suggestion of Schmidt (2020). The Area Deprivation Index assigns each census block group to a decile of relative disadvantage by taking a weighted combination of various measures of disadvantage (Kind and Buckingham 2018) and ranking all census block groups nationally. Unlike directly using race or ethnicity, which may be subject to court challenges, ADI does not directly condition on race or ethnicity, but on neighborhood characteristics like education, income, and homeownership rates. There is precedent for using neighborhood proxies in rationing, for the context of school seat assignment. In the 2007 Parents Involved in Community Schools vs. Seattle case, the United States Supreme Court decision ruled against directly using race as a factor in K-12 admissions. Starting in 2009, Chicago Public Schools adopted a neighborhood-based affirmative action system, using only census tract level measures of disadvantage (Dur, Pathak and Sönmez 2020; Ellison and Pathak 2020).

ADI has also been used in recent weighted lottery-based schemes, a special case of reserve systems, to allocate anti-viral drugs (Pathak et. al. 2020b). The Commonwealth of Pennsylvania has used ADI as part of the state's official policy for allocating scarce COVID-specific medications, including Remdesivir. COVID patients who reside within the most disadvantaged neighborhoods in Pennsylvania receive higher priority weights to receive these scarce medications as part of the State's allocation lottery (Pennsylvania 2020). An identical system was first deployed as part of the University of Pittsburgh Medical Center's triage protocol design for Remdesivir. In both systems, if a neighborhood has an ADI score of 8 or greater (on a scale of 1-10), patients receive heightened priority for the anti-viral drug (White et al. 2020). For pragmatic reasons, and we follow this precedent and refer to an ADI score of 8 or greater as a high ADI.

Figure 3 reports on the proportion of vaccine doses allocated to high ADI areas under the 2018 CDC Vaccine allocation guideline. ADI is calculated in deciles, so a score of 8 or higher corresponds to 30% of US neighborhoods. The Figure illustrates that high ADI neighborhoods receive less than 30% of vaccines if the total supply of vaccines is less than 26 million. That is, if there are not even enough units to allocate to first tier of patients, then high ADI areas will be underrepresented. However, as soon as the number of units is enough for Tier 2, roughly 30% of units will be allocated to high ADI areas, a fair share relative to their geographic representation.

To use ADI as part of a vaccine rationing scheme, we illustrate a reserve system with two possibilities: either a 20% or 40% high-ADI reserve, implemented in an over-and-above way. This means that for a given total number of vaccine units, we categorize them in two groups. The first group is offered vaccine doses based on the as-is, unmodified 2018 CDC priority system. After allocating those units (it is expected that the first tranche will be fully exhausted), the system then allocates the high-ADI reserve units, by giving any individual in a high ADI area the highest priority. If all high-ADI individuals receive a vaccine, then any remaining units in the high-ADI reserve will be awarded based on the 2018 CDC priority system. In other words, once all high-ADI individuals receive their vaccines, the overall system reverts back to the 2018 CDC priority system.

The allocation system we consider only uses the high-ADI reserve after allocating all units to individuals in Tier 1. That is, after all units are allocated to individuals in Tier 1 according to the 2018 CDC vaccine guidelines, any remaining units are then categorized according to the specified reserve size. We introduce the reserve only after Tier 1 because Tier 1 covers essential homeland and national security personnel, critical healthcare workers, law enforcement personnel, pregnant women, and infants and toddlers. We believe that if there are not enough vaccines to even reach this initial group, considerations associated with targeting the hardest-hit would be subject to increased scrutiny.

Figure 4 illustrates how a reserve system would work for vaccines with a high-ADI reserve category in addition to the original 2018 CDC category. The top panel shows the proportion of incremental vaccine units awarded to the high ADI population as the total supply of doses increases. When there are fewer than 5 million doses, no more than 25% of units are awarded to the high ADI population. However, as the number of doses increases, the marginal share jumps to nearly 40%. This results in the cumulative share of vaccines going to high-ADI areas being 30% when there are enough doses to accommodate all individuals in Tier 1, shown in Figure 3.

The second panel of Figure 4 illustrates how the 20% high-ADI reserve works. Since the reserve system only applies after units are allocated to al Tier 1 recipients, the first change in the marginal share is at Tier 2. At that point, the marginal share of high-ADI population receiving a vaccine is roughly 45%, compared to 30% in the absence of any high-ADI reserve given in the top panel. With a 20% high-ADI reserve, the fraction jumps because a fifth of vaccines are awarded to the high ADI areas on top of the existing units (80% of 30%) they obtain without such a reservation. Therefore, the high-ADI reserve provides an additional boost that amounts to 20% for the reserve in addition to the $0.8 \times 30\% = 24\%$ through the unreserved units, for a total of 44%.

The second panel also shows how a high-ADI reserve works by placing high ADI individuals ahead of where they would be prioritized without a high-ADI reserve. When the total number of vaccines is 230 million, there are no remaining high-ADI individuals left and each has been allocated a vaccine.

The third panel repeats the same calculation for a 40% high-ADI reserve. As before, the reserve applies only after units are allocated to individuals in Tier 1. At that point, the marginal share of units allocated to high ADI jumps to nearly 60%. (As before, $40\% + 0.6 \times 30\% = 58\%$). With a higher reserve size, the number of high-ADI individuals without a vaccine is exhausted by 175 million person-unit vaccines.

### B. High-ADI Reserve in Action

A high-ADI reserve results an increase in the proportion of vaccines allocated to Black and Indigenous recipients. This is shown in Figure 5, which plots the combined Black and Indigenous allocation for a 20% and 40% high-ADI reserve as well as the 2018 CDC guidelines. The ADI increases the proportion of vaccines allocated to Black and Indigenous, with a peak when there are about 75 million units of a vaccine. At that level of supply, a 20% high-ADI reserve results in 20% of vaccines allocated to Black and Indigenous and a 40% high-ADI reserve results in about 22% of vaccines allocated to them. Both fractions are still well below the proportion of COVID deaths or age-adjusted deaths for Black and Indigenous communities.

While the ADI results in an increase in the share of vaccines allocated to Black and Indigenous persons, it has a modest effect on the regional distribution of vaccines. This can be seen in Figure 6, which shows how the share calculated either relative to a state's share of population, share of cases, or share of deaths changes with a high-ADI reserve. Relative to the state share, a high-ADI reserve increases the representation of several southern states, but by a modest amount. The effect on the other two measures of fair shares due to a high-ADI reserve is negligible.

The Appendix reports on several other dimensions of vaccine allocation. Figure A1 adds Hispanic populations to Figure 5 and shows that a high-ADI reserve results in a higher share of Black, Indigenous, and Hispanic recipients. At 75 million units, a 40% high-ADI reserve results in more vaccines to this group than both their share of the population and their share of COVID deaths. Figure A2 compares a high-ADI reserve to a direct racial reserve for Black and Indigenous individuals. For levels of vaccines under 100 million person units, a 40% high-ADI reserve has a similar effect as a 5% reserve for Black and Indigenous individuals. Figure A3 shows that the average age of vaccine recipients is lower than the population average under 2018 CDC guidelines for most levels of vaccine supply. Figure A4 shows that the share of vaccines allocated to females is higher than the population fraction for vaccine supply less than 100 million person units, but is close to the population female fraction when supply is greater than 100 million person units.

## V. Conclusion

The current debate on how to prioritize vaccine allocation for hardest-hit communities has not wrestled with specific details of whether existing systems are equitable based on empirical data or how to actually implement heightened priority for particular groups. Since major groups have not yet finalized guidelines on allocation despite work, we report on how the 2018 CDC Pandemic Influenza Vaccine guidelines would allocate vaccines to Black and Indigenous individuals and across states.

Our analysis illustrates the importance of a precise definition of equitable allocation in setting vaccine allocation policy. We estimate that Black and Indigenous populations combined receive more vaccines compared to their total population share for all levels of vaccine supply. However, they receive a substantially lower share compared to their share of COVID deaths or age-adjusted deaths. Moreover, whether the CDC guidelines provide an equitable distribution of vaccines across states depends crucially on whether population, case-count, deaths, or other reference points are the appropriate benchmark.

This paper also reports on an implementation of a vaccine allocation system targeting the hardest-hit communities using a reserve system. In our illustrative reserve system, individuals from neighborhoods with a high ADI receive preferential treatment for a subset of the total vaccine supply. A reserve system offers flexibility over a priority system because not all units have to be assigned according to the same priority order. A 40% high-ADI reserve increases the share of vaccines to Black and Indigenous individuals, with a modest effect on the distribution of vaccines across states. A larger reserve size would increase the share for Blacks and Indigenous individuals further. There are many levers in a reserve system we could consider beyond the reserve size. However, a reserve system offers a feasible way to implement heightened priority for disadvantaged groups and has more flexibility than a priority system.

Aside from alternative parameters in a reserve system, there are several directions for further study. First, our estimate required extrapolation from ACS and BRFSS data, and current estimates of COVID cases and deaths. As improved data become available, it would be worthwhile to revisit these estimates. The second issue involves take-up. Several studies suggest that Black and Hispanic (both Hispanic-black and Hispanic-white) adults are among the least likely to voluntarily take vaccines compared to whites and Asians (Lu et al. 2014, Quinn et al. 2017). If the goal is to protect the health of these groups, then the system must consider who is allocated the declined unit, once reasonable efforts to convey the benefits of vaccinations have been exhausted. Likewise, there may be instances where a high priority individual does not take a vaccine and instead wishes to give her vaccine access to a lower priority individual, such as a family member. These kinds of situations may alter the overall distribution of vaccine recipients. Finally, our analysis has focused on the case where there is a single type of vaccine that is distributed nationally. New complications and opportunities arise when distribution is not nationally coordinated and there are multiple vaccines that differ in their effectiveness, safety, or cost.

**Vaccination tiers and population groups for a high/very high level of pandemic severity.**

Accessible version at https://www.cdc.gov/flu/pandemic-resources/national-strategy/planning-guidance/guidance_508.html#figure-1

*This figure illustrates how vaccination is administered to population groups by tiers until the entire U.S. population has had the opportunity to be vaccinated during a pandemic with a high or very high level of severity, and how tiers integrate population groups, balancing vaccine allocation to occupationally defined groups and the general population. (See Appendix A for description of Occupational and High Risk Population Groups)*

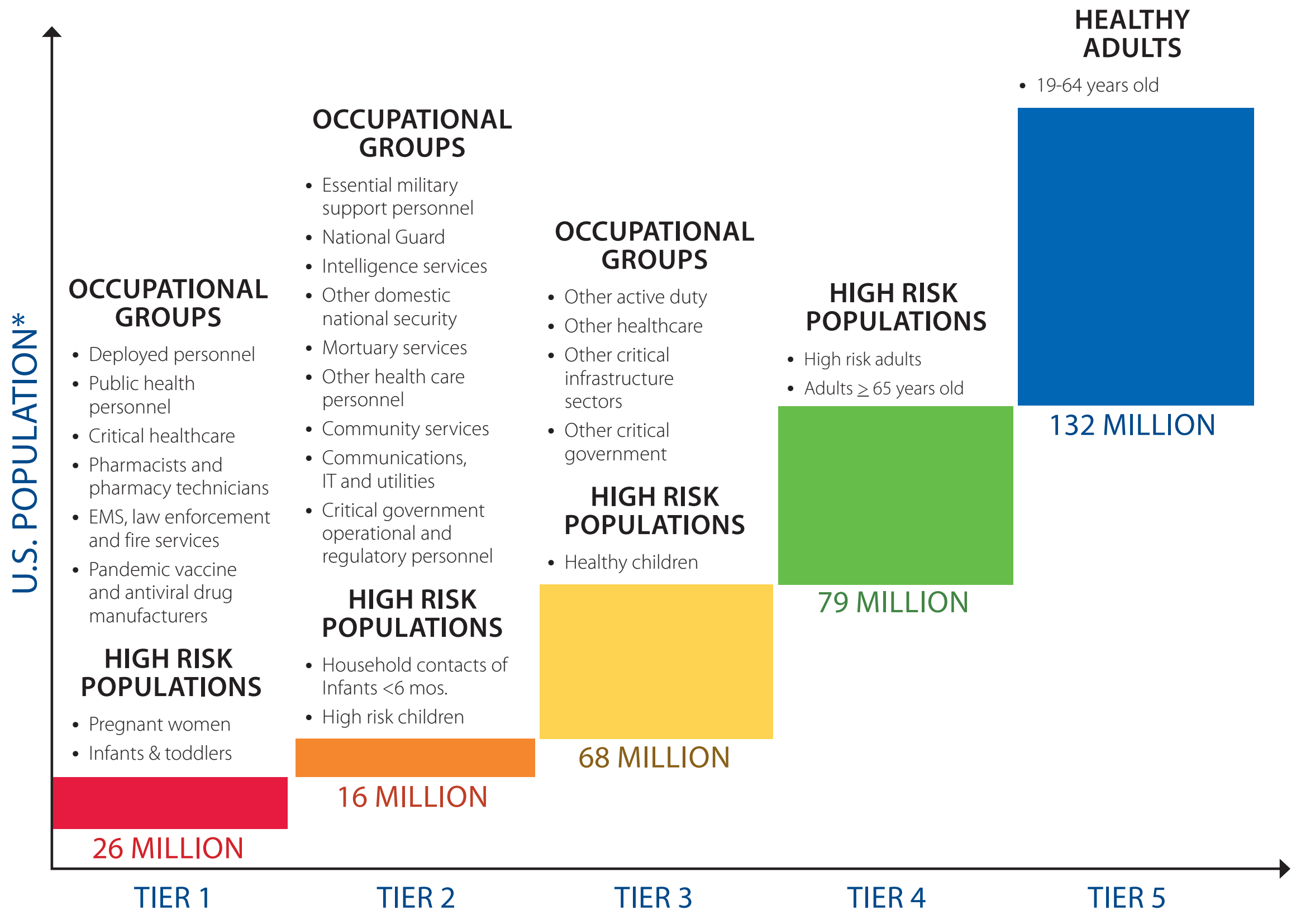

* Based on 2015 U.S. population of 321 million people. https://factfinder.census.gov/bkmk/table/1.0/en/PEP/2015/PEPAGESEX

**Table 1. Vaccination Tiers in 2018 CDC Guidelines**
(Available at: https://www.cdc.gov/flu/pandemic-resources/pdf/2018-Influenza-Guidance.pdf)

# Figure 1
## Black + Indigenous population

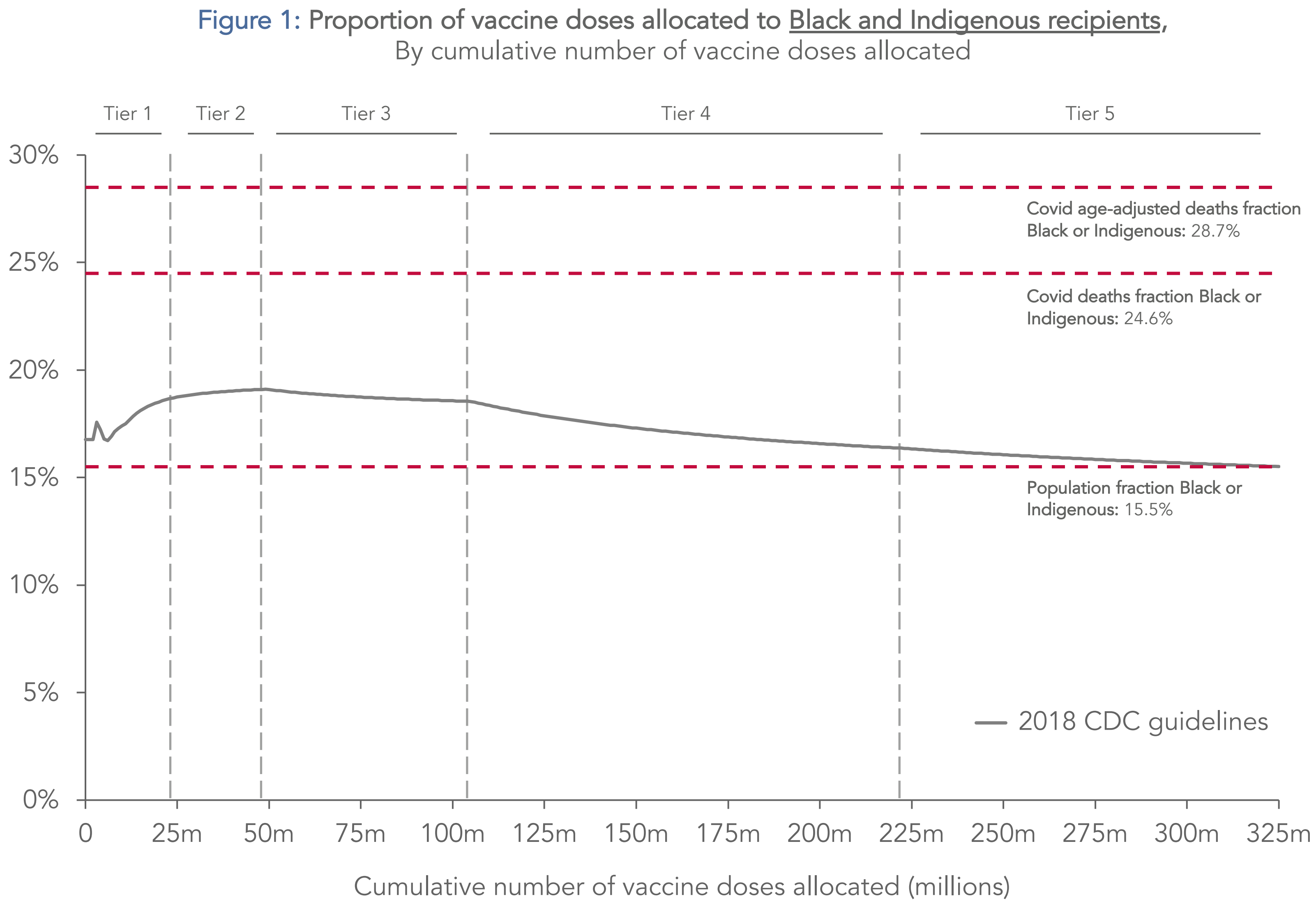

Note: Death rates may be inaccurate in the presence of multi-racial groups and racial groups outside of Indigenous, Asian, Black, Latino, White, and Pacific Islander
Source: CDC Vaccine Allocation Guidelines; American Community Survey (ACS) 5-Year; APM Research Lab, Color of Coronavirus

## Figure 2: State share of vaccines

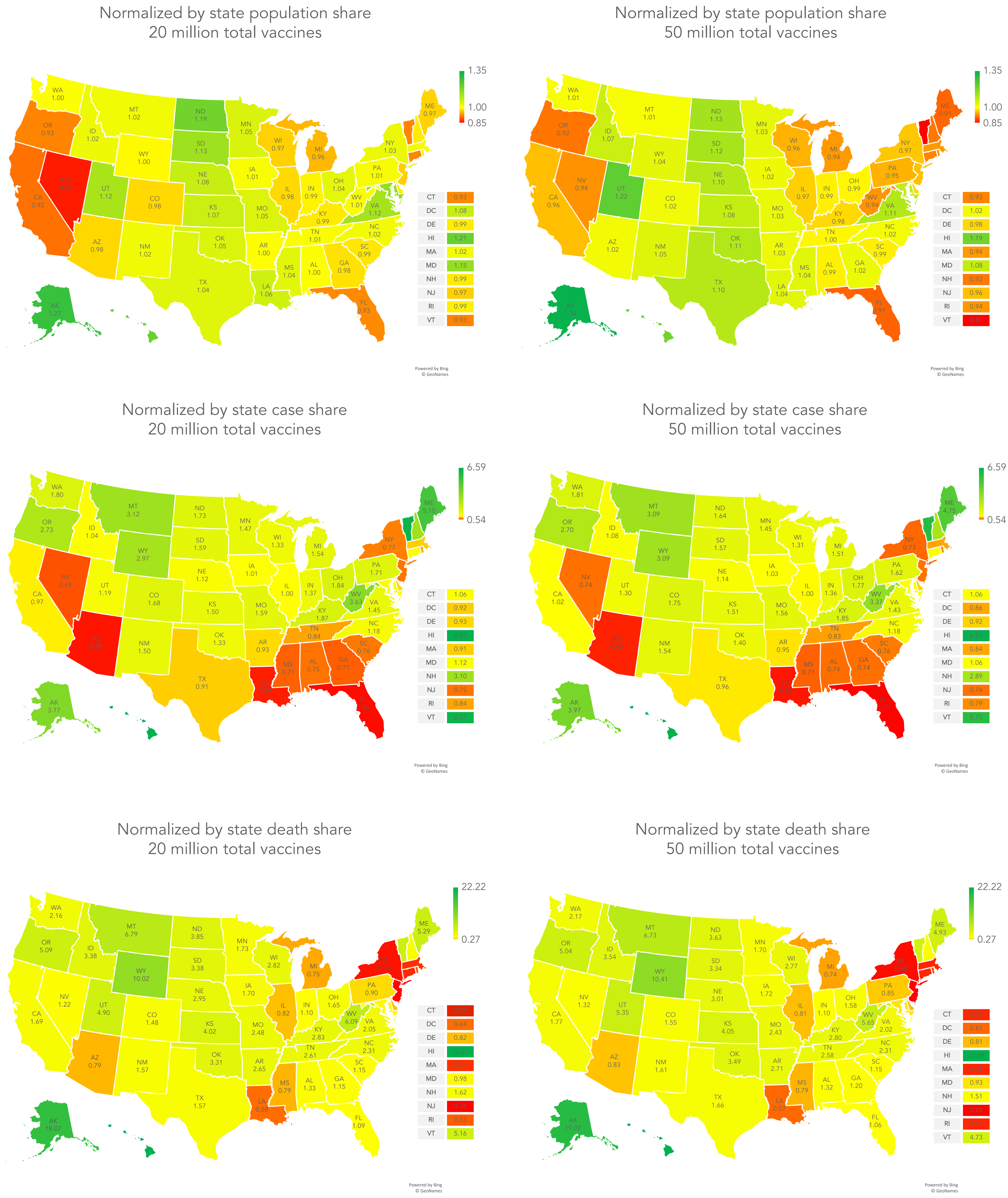

Note: Area Deprivation Index (ADI) is based on measure created by the Health Resources & Services Administration (HRSA) to allow for rankings of neighborhoods by socioeconomic disadvantage in a region of interest
Source: CDC Vaccine Allocation Guidelines; American Community Survey (ACS) 5-Year; APM Research Lab, Color of Coronavirus

# Figure 3
## High ADI population - proportion

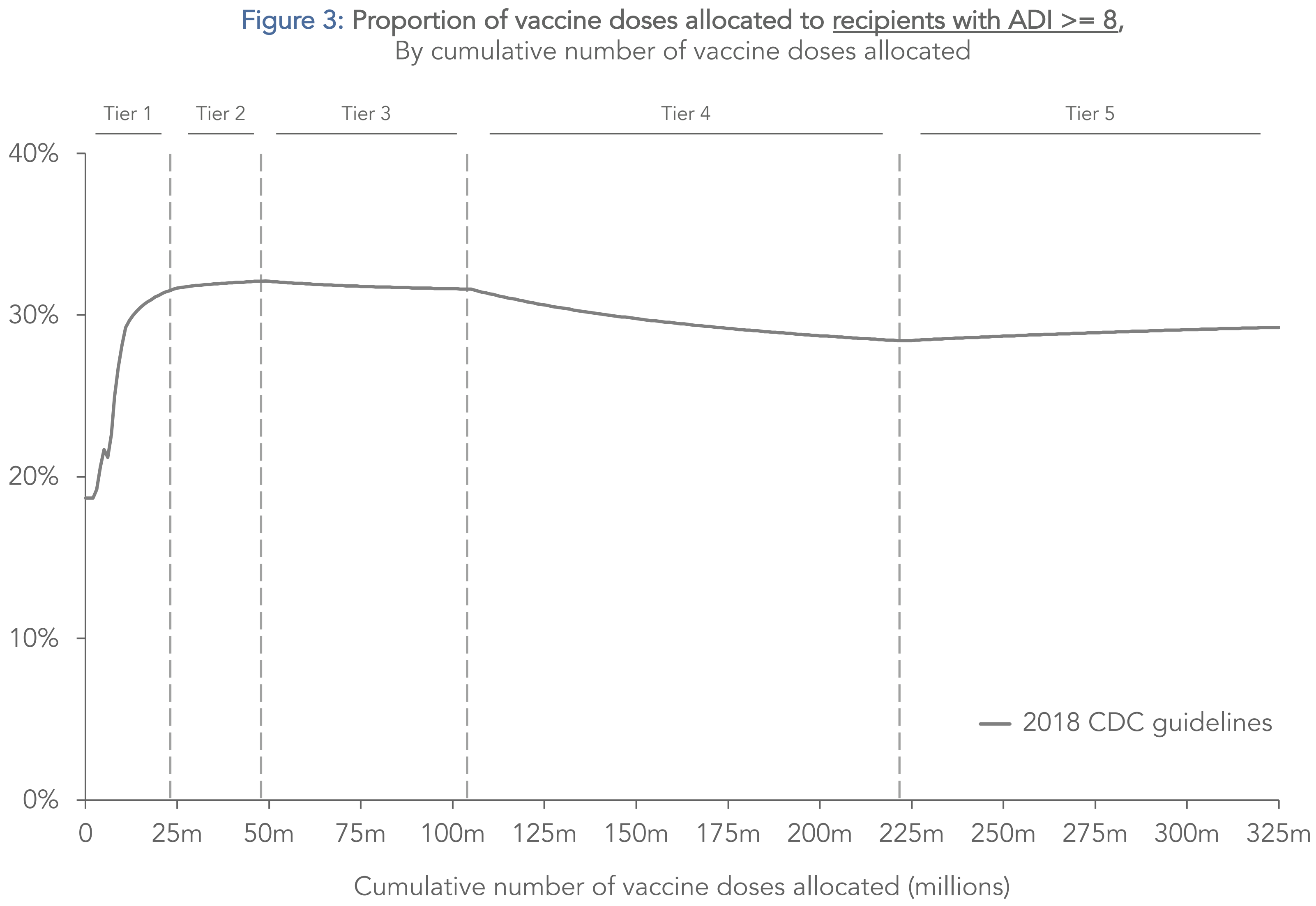

Note: Area Deprivation Index (ADI) is based on measure created by the Health Resources & Services Administration (HRSA) to allow for rankings of neighborhoods by socioeconomic disadvantage in a region of interest
Source: CDC Vaccine Allocation Guidelines; American Community Survey (ACS) 5-Year; APM Research Lab, Color of Coronavirus

# Figure 4
## Marginal share of high ADI population obtaining a vaccine

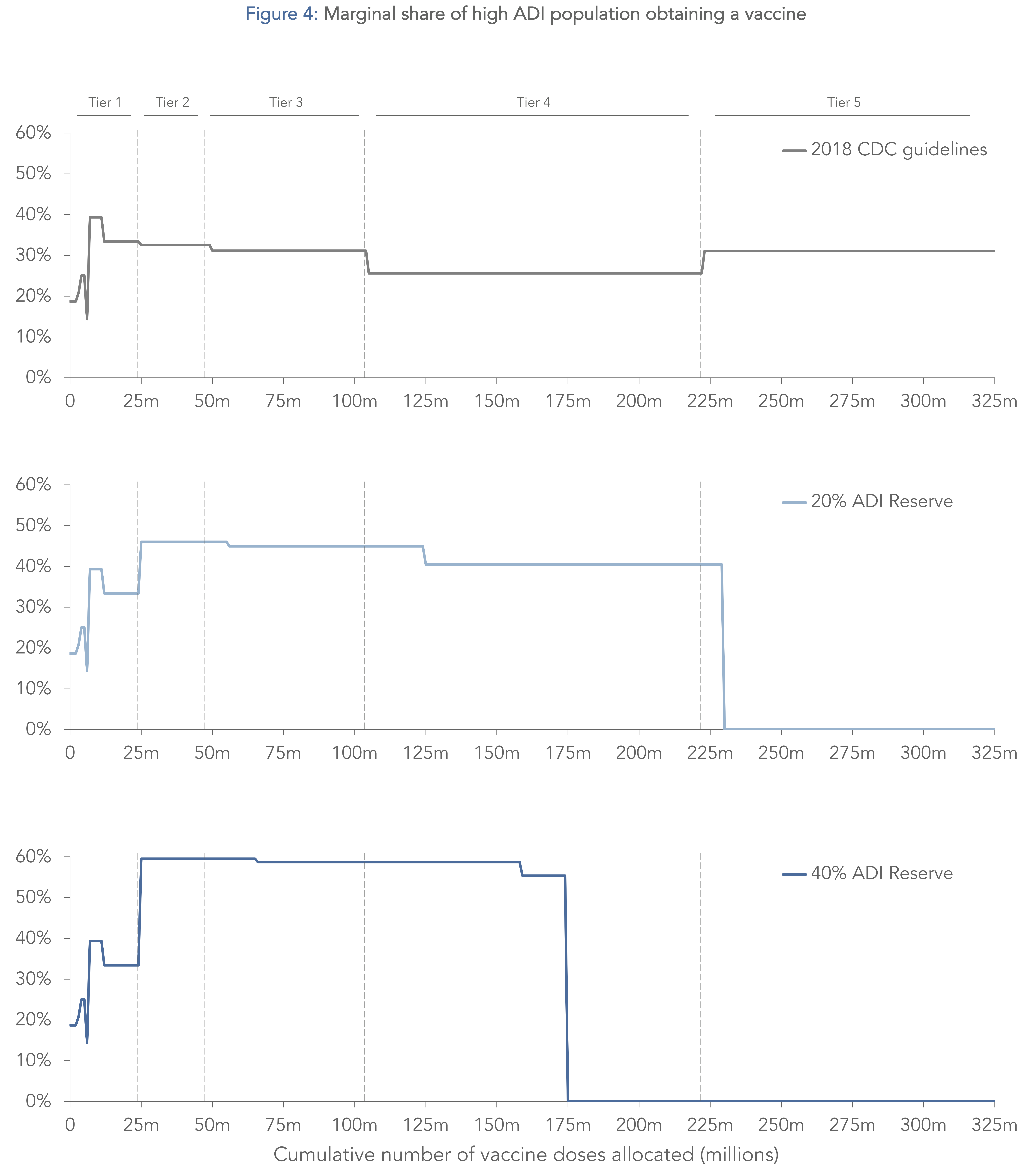

Note: Area Deprivation Index (ADI) is based on measure created by the Health Resources & Services Administration (HRSA) to allow for rankings of neighborhoods by socioeconomic disadvantage in a region of interest
Source: CDC Vaccine Allocation Guidelines; American Community Survey (ACS) 5-Year; APM Research Lab, Color of Coronavirus

# Figure 5
Black + Indigenous population

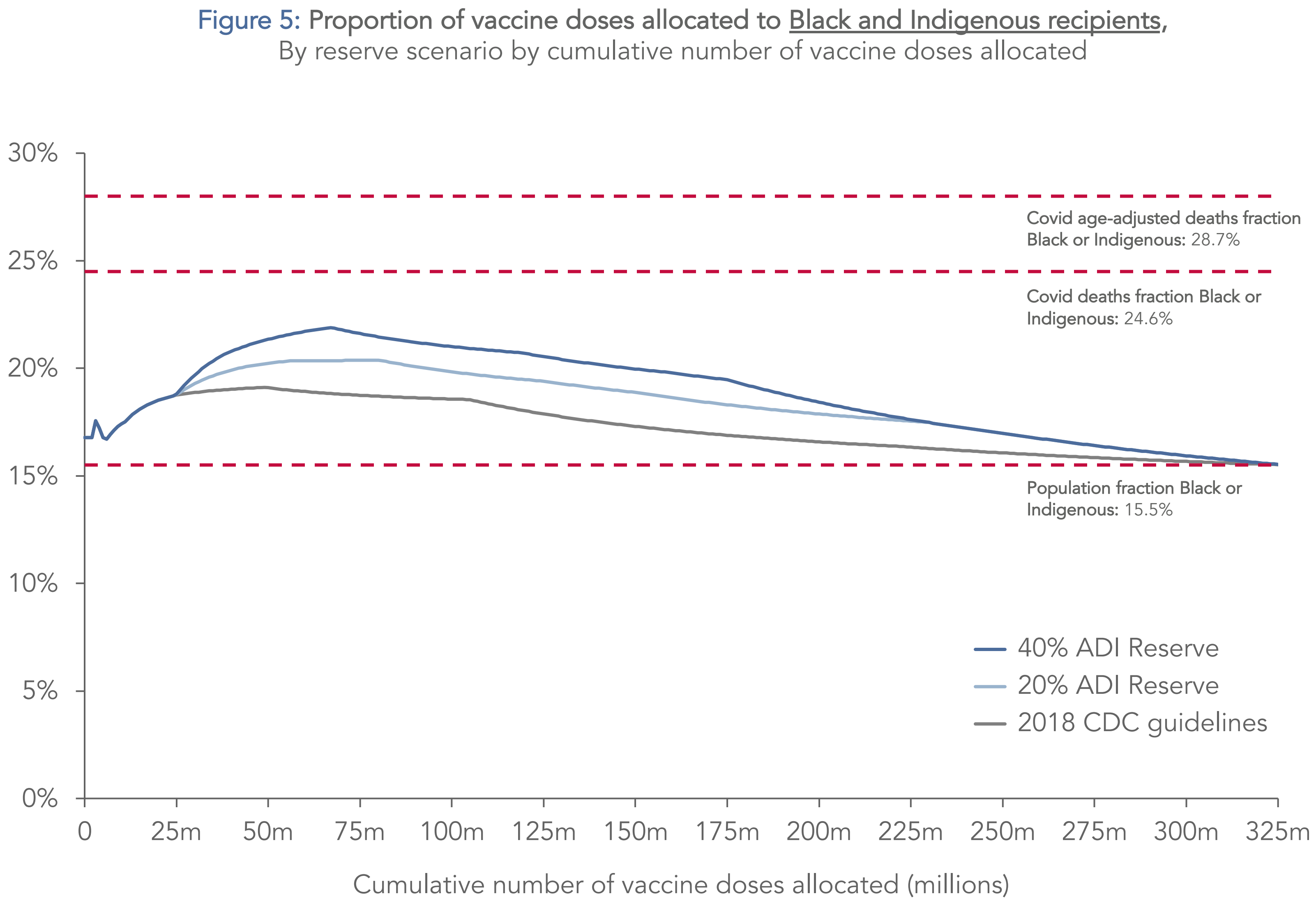

Note: Area Deprivation Index (ADI) is based on measure created by the Health Resources & Services Administration (HRSA) to allow for rankings of neighborhoods by socioeconomic disadvantage in a region of interest; death rates may be inaccurate in the presence of multi-racial groups and racial groups outside of Indigenous, Asian, Black, Latino, White, and Pacific Islander
Source: CDC Vaccine Allocation Guidelines; American Community Survey (ACS) 5-Year; APM Research Lab, Color of Coronavirus

Figure 6: Vaccines per state as proportion of "fair share" for 50 million total vaccines

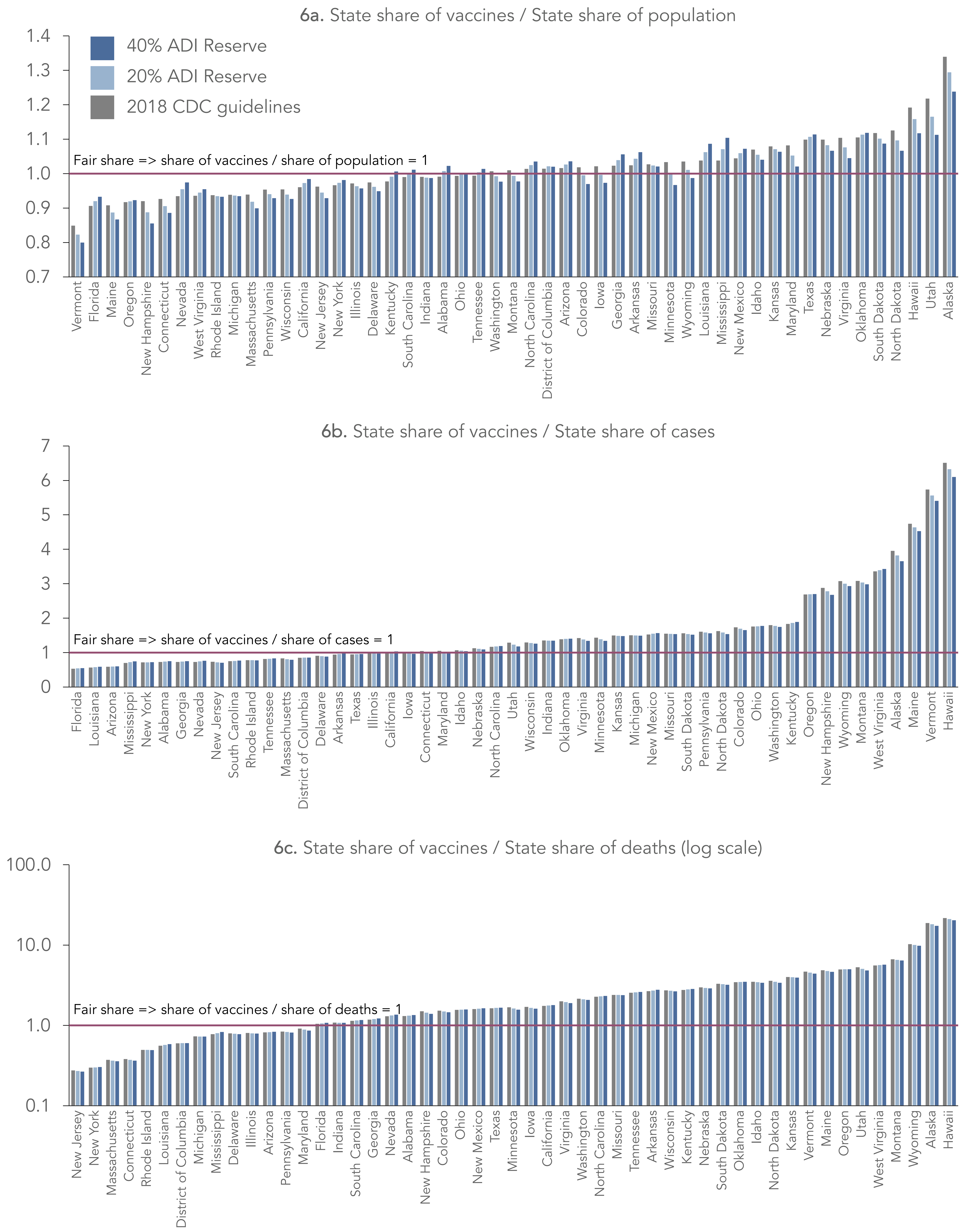

Note: Area Deprivation Index (ADI) is based on measure created by the Health Resources & Services Administration (HRSA) to allow for rankings of neighborhoods by socioeconomic disadvantage in a region of interest

Source: CDC Vaccine Allocation Guidelines; American Community Survey (ACS) 5-Year; APM Research Lab, Color of Coronavirus

# Figure A1
Black + Indigenous + Hispanic population

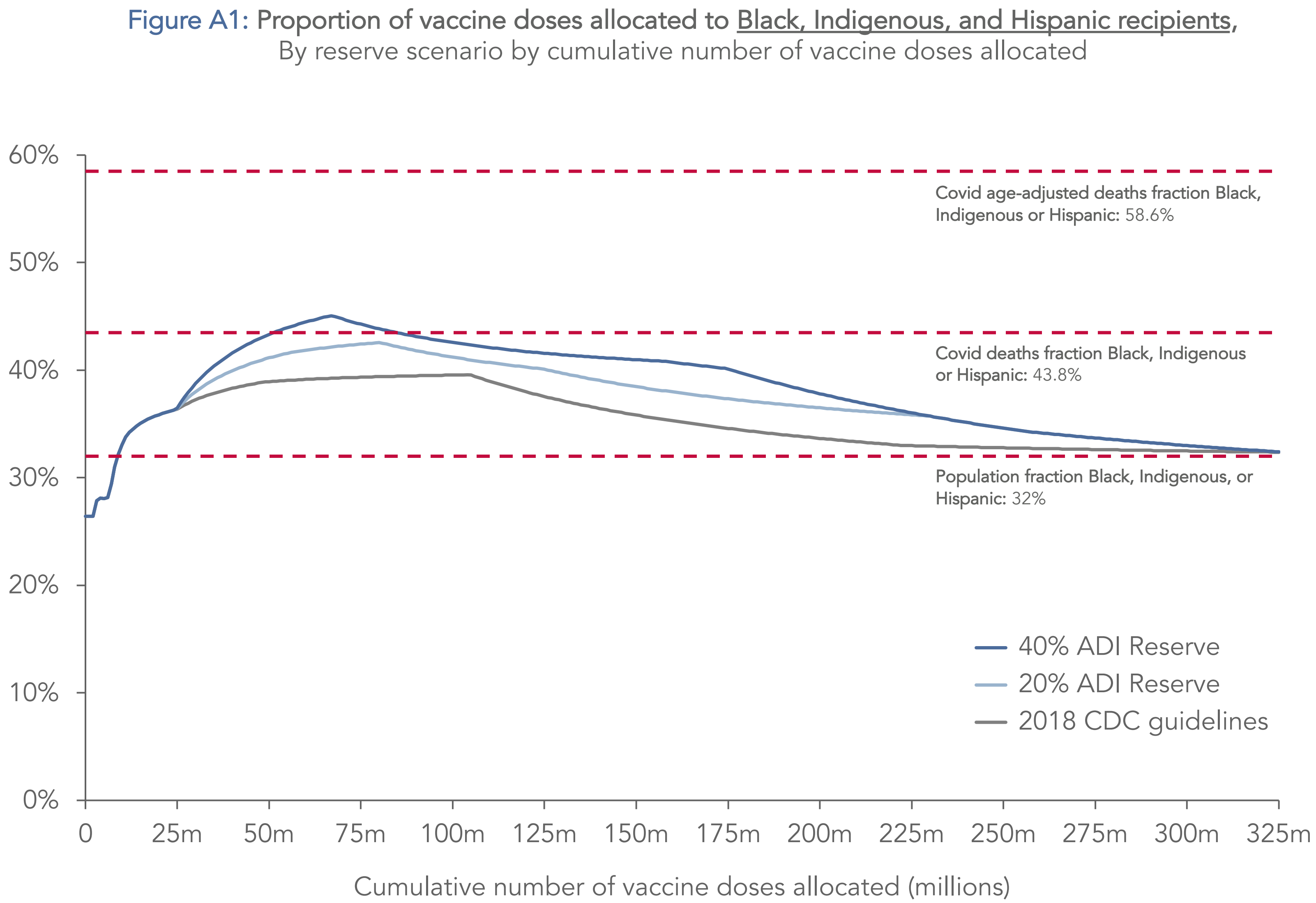

Note: Area Deprivation Index (ADI) is based on measure created by the Health Resources & Services Administration (HRSA) to allow for rankings of neighborhoods by socioeconomic disadvantage in a region of interest; death rates may be inaccurate in the presence of multi-racial groups and racial groups outside of Indigenous, Asian, Black, Latino, White, and Pacific Islander
Source: CDC Vaccine Allocation Guidelines; American Community Survey (ACS) 5-Year; APM Research Lab, Color of Coronavirus

# Figure A2
## Black + Indigenous proportion under ADI reserve vs. Black + Indigenous reserve

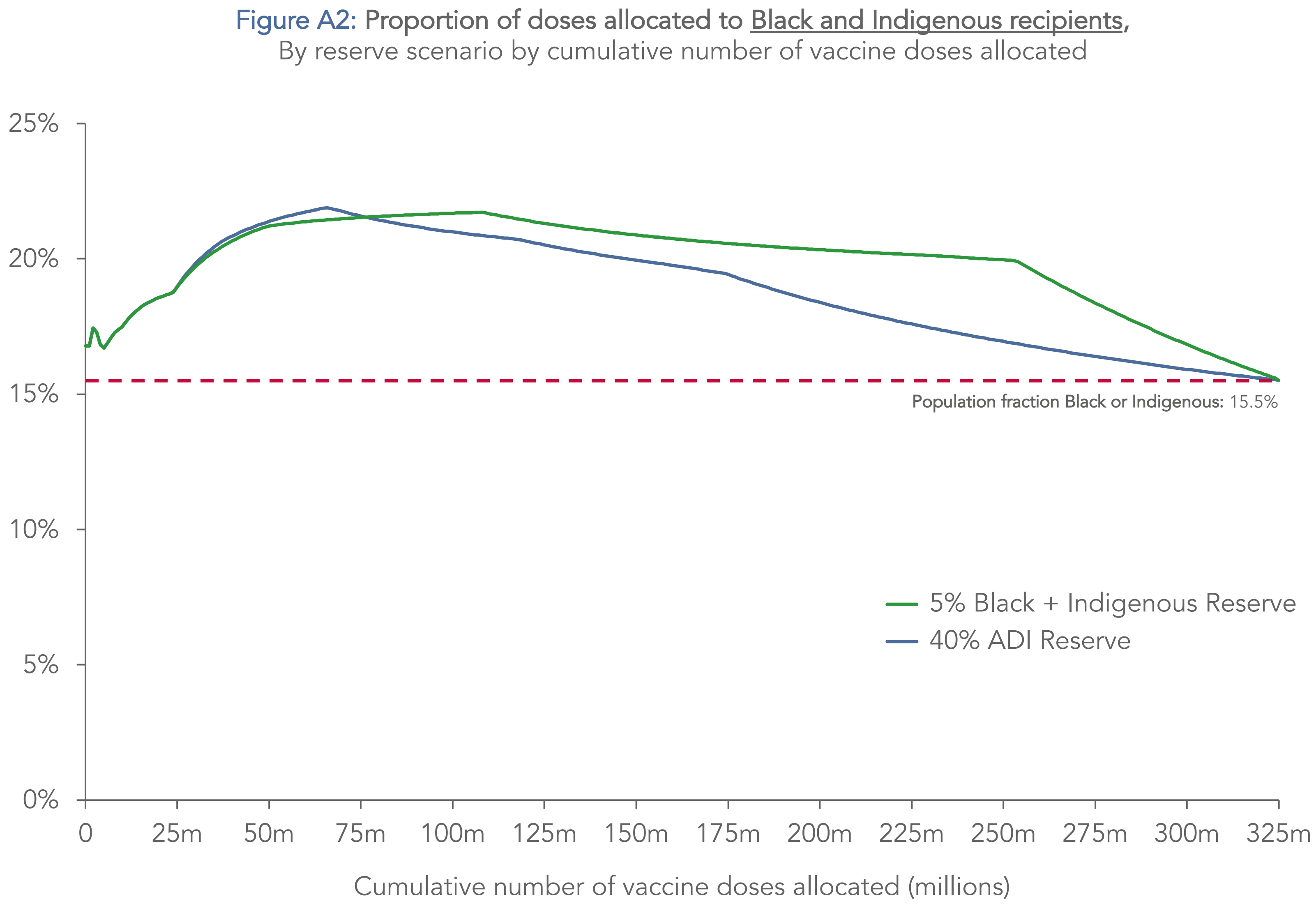

Note: Area Deprivation Index (ADI) is based on measure created by the Health Resources & Services Administration (HRSA) to allow for rankings of neighborhoods by socioeconomic disadvantage in a region of interest
Source: CDC Vaccine Allocation Guidelines; American Community Survey (ACS) 5-Year; APM Research Lab, Color of Coronavirus

# Figure A3
## Average age

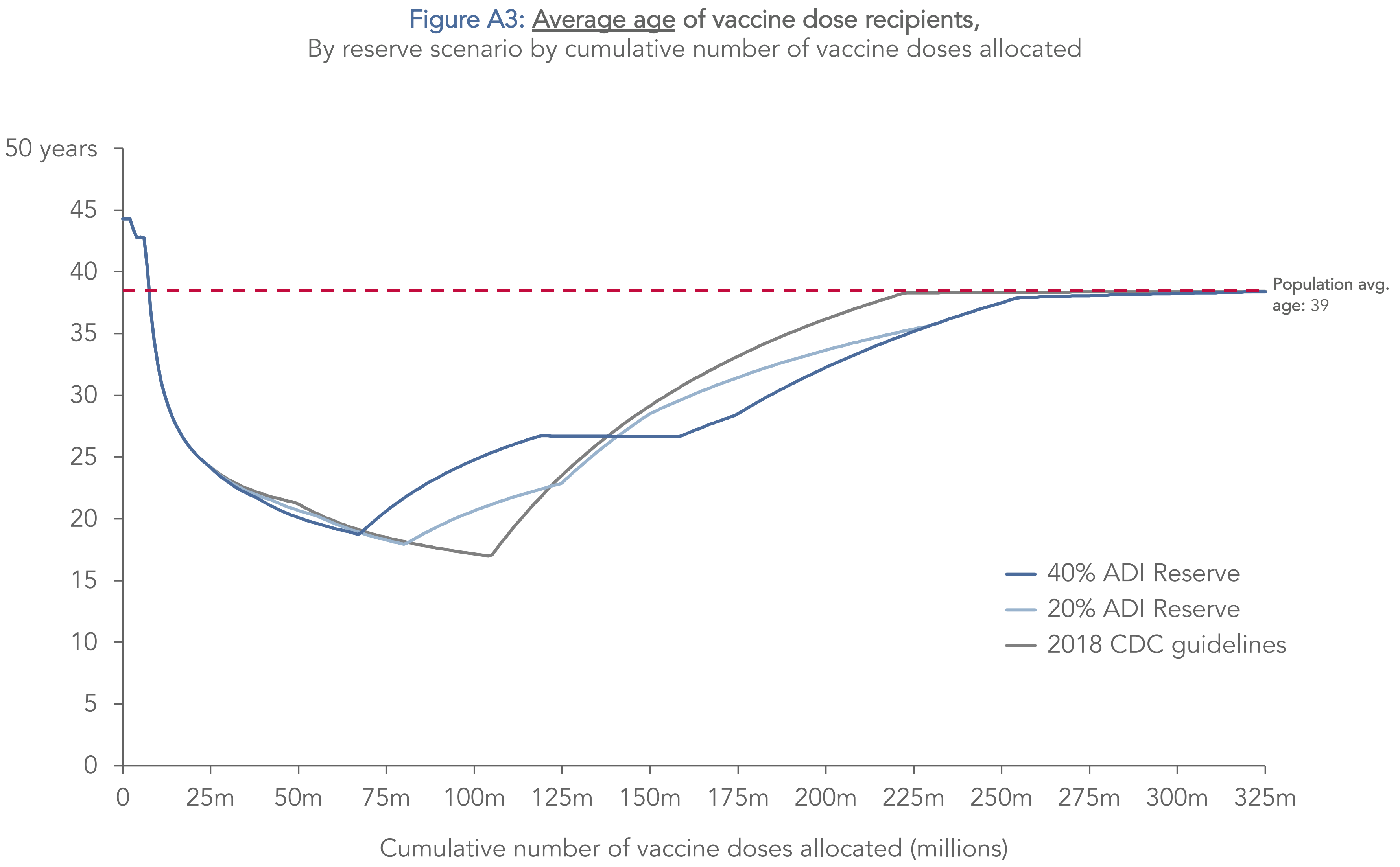

Note: Area Deprivation Index (ADI) is based on measure created by the Health Resources & Services Administration (HRSA) to allow for rankings of neighborhoods by socioeconomic disadvantage in a region of interest
Source: CDC Vaccine Allocation Guidelines; American Community Survey (ACS) 5-Year; APM Research Lab, Color of Coronavirus

# Figure A4
## Female population

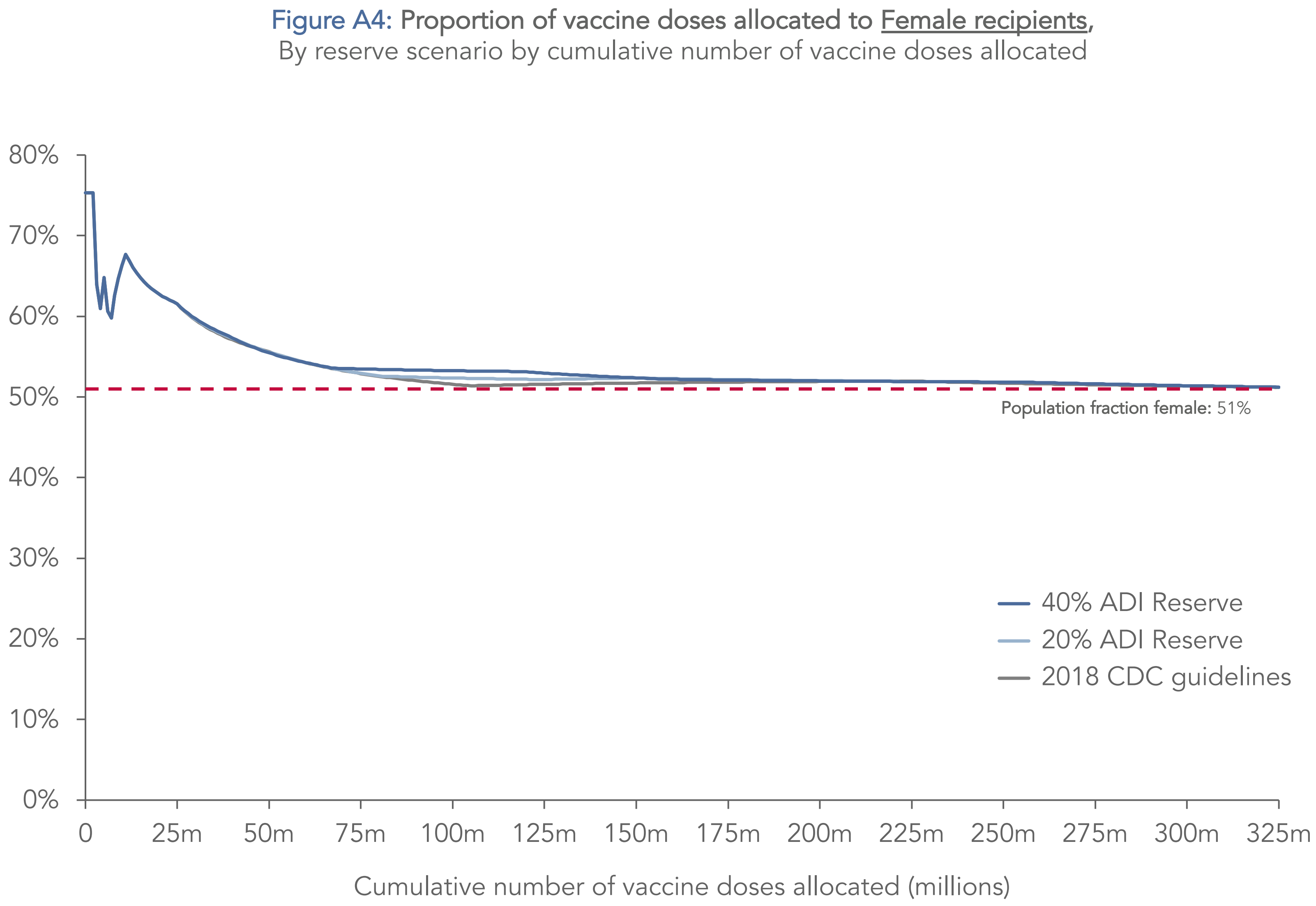

Note: Area Deprivation Index (ADI) is based on measure created by the Health Resources & Services Administration (HRSA) to allow for rankings of neighborhoods by socioeconomic disadvantage in a region of interest
Source: CDC Vaccine Allocation Guidelines; American Community Survey (ACS) 5-Year; APM Research Lab, Color of Coronavirus

**Online Appendix: Simulation Procedure and Data Sources**

to "Do Black and Indigenous Communities Receive their Fair Share of Vaccines Under the 2018 CDC Guidelines?" By

Parag A. Pathak, Harald Schmidt, Adam Solomon, Edwin Song, Tayfun Sönmez, M. Utku Ünver

**Contents**

## A1. Overview

### A1.1. Data Sources

We base our priority tiers off of the priority tiers for a severe pandemic outlined in the 2018 CDC guidance (CDC 2018a) on vaccine prioritization during an influenza pandemic. The US population is partitioned into 5 tiers based on essential or high-risk occupations and demographic factors indicating risk for spreading/contracting severe disease. In case there are not enough vaccines to go around for even the first tier, the first tier is further partitioned into 7 subtiers. For tiers 2-5, those in the same tier have equal priority. We assume that among individuals with the same priority, the vaccine will be distributed at random.

The 2018 CDC document contains estimates, to the nearest 50k, of the number of people within each group within each tier. These numbers come from "information provided by Department of Defense, Department of Homeland Security, Department of Health and Human Services, and U.S. Census Bureau." These priority tiers sometimes overlap. If someone qualifies for two or more tiers, they are vaccinated with the highest tier they qualify for. However, the document and surrounding CDC sites do not specify whether they would be included in each tier's estimated numbers or only that of the highest tier. In our process, we are assuming that the CDC numbers encompass those who qualify for higher tiers as well.

As the CDC's data is unavailable to us, we use the American Community Survey (ACS) 2014-2018 5-year Public Use Microdata Sample (PUMS) (U.S. Census Bureau 2014-2018). The 5-year PUMS contains two datasets, one containing housing unit characteristics for a sample of housing units and another containing individual characteristics for the individuals within those same housing units. We use the person data, which has over 15 million observations, for our labelling of tiers.

The PUMS person data is a weighted sample. Every observation of an individual in the PUMS person data is associated with a weight called "person's weight for generating statistics on individuals" (PWGTP). This weight is "used to bring the characteristics of the sample more into agreement with those of the full population by compensating for differences in sampling rates across areas, differences between the full sample and the interviewed sample, and differences between the sample and independent estimates of basic demographic characteristic estimates of population characteristics", according to the ACS Design and Methodology report (U.S. Census Bureau 2014).

PWGTP was also scaled such that the size of any population group could be estimated by the sum of PWGTP across observations in the PUMS belonging to the group. Further details about calculation of weights in the ACS can be found in the PUMS technical documentation (U.S. Census Bureau 2020a).

In our tier labelling, we have assumed that the CDC-provided numbers include those who may also qualify for a higher tier, unless the CDC-provided group definitions say otherwise. Our procedures allow each observation to be marked for as many tiers as the observation qualifies for. There are specific pairs/sets of groups within tiers that cannot overlap, either by the CDC's explicit statement ("not included in higher priority groups"), the CDC's implicit statement ("other groups"), or common sense (two groups that the PUMS doesn't differentiate but should not actually overlap). Our understanding of those restrictions has been accounted for in our labelling criteria.

Due to data limitations, we rely on randomization for the assignment of people to certain CDC groups.

The ACS PUMS does not include data on the presence of high-risk conditions, which impacts one's tier priority. We supplement with 2018 data from the Behavioral Risk Factor Surveillance System (BRFSS) (CDC 2018b). As the observations in the BRFSS are unrelated to those in the PUMS, we rely on randomization to link the two datasets by computing the proportion with a high-risk condition for each characteristic demographic group in the BRFSS and assigning each observation in PUMS person data as high-risk using weighted coin flips. More details are in the "Tier inclusion criteria" section of this document.

We chose to run the tier assignment process only once due to the large size of the dataset and the large sizes of groups and demographic categories relative to PWGTP. This is not a Monte Carlo simulation.

We use both the person and the housing unit PUMS data for assigning Area Deprivation Index (ADI) 2.0 (University of Wisconsin School of Medicine and Public Health 2015) to individuals, as the components of the ADI involve both individual-level attributes found only in the person data and household-level attributes found only in the housing unit data. We were able to link the two datasets, as well as identify residents of the same housing unit within the person data, using the variable SERIALNO, an ID number unique to the housing unit that is present in both the person and the household data.

We cannot assign the actual ADI of each person in the PUMS data because ADI is computed using census block level averages and the PUMS geographic specificity only goes down to the Public Use Microdata Area (PUMA) level. We instead approximate the ADI of each family using averages taken at the family level. We then use SERIALNO to assign to each individual in the person data the ADI of her family We do not perform any randomization in computing ADI. Data on proportion of COVID deaths by race/ethnicity and age-adjusted proportion of COVID deaths by race/ethnicity are sourced from APM Research Lab (2020). The data was accessed August 11, 2020 and released August 5, 2020, reflecting data releases through August 4.

### A1.2. Simulation Procedure

We consider three treatment groups for strict CDC guidelines, a reserve system with 20% over-and-above high-ADI reserve, and a reserve system with 20% over-and-above high-ADI reserve while the units within the unreserved category and reserve category, including the unreserved category are allocated using the CDC guidelines. We follow the following procedure in our simulation analysis.

1. Approximate the ADI of each household as explained in "Calculation of ADI Values" section of this appendix.

2. Label every person in the PUMS with every group they qualify for. PUMS variables do not perfectly match up to the groups but can be used as a crude approximation. Our inclusion criteria for each tier is outlined in the "Tier inclusion criteria" section of this appendix.
3. Determine the highest-ranking tier for which each person qualifies. Determine, also, whether each person has an ADI in the top 30% of people in the nation we calculate an ADI for
4. For each tier, calculate the demographic averages (proportion breakdown of race, gender, age, etc.) and total populations of those in the PUMS who qualify for that tier but do not qualify for any higher tier.
5. Changing total vaccine quantity in 10,000-unit increments of total vaccines from 0 to 100,000 and then in increments of 100,000 vaccine units from 100,000 to 325.7 million (the US population plus duplicated entries for pregnant women, see the A2.2.4), for each total vaccine quantity, do the following:
   5.1. Using the tier and ADI priority and group size, calculate how many of each group would receive a vaccine for each amount of total vaccine units under the CDC guidelines and the reserve system for each high-ADI reserve category size. Assume that for those of the same priority, vaccine distribution is randomized.
   5.2. Using the PUMS data, calculate the proportion Black or Indigenous; proportion Black, Indigenous, or Latino/Hispanic; proportion female; and average age of individuals who are allocated vaccines under the CDC guidelines and the reserve system for 20% and 40% high-ADI reserve category.
6. For each state, estimate state population using PUMS individual-level data set, and number of deaths and cases for each state using CDC data retrieved August 14 (CDC 2020b).
7. For two total vaccine quantities, 20 million and 50 million, do the following:
   7.1. Using data vaccine allocation shares found in Step 5, and population estimates found in Step 6, calculate allocation for each state under the CDC guidelines and the reserve system for 20% and 40% high-ADI reserve category size. Note that for 20 million vaccines, the allocation under the CDC guidelines and both reserve sizes are the same.
   7.2. Calculate "fair share" indices relative to population, deaths, and cases using the estimates per state under the CDC guidelines and the reserve system for 20% and 40% high-ADI reserve category size.

Details about graphing are summarized in the last section of this appendix.

We use statistical open-source R software (version 3.6.1) and Microsoft Excel in our simulation. the names of program files are given in a table at the end of this appendix. The programs and data files are available from the authors upon request.

## A2. Tier Labelling Criteria

### A2.1. CDC groups and tiers

We integrate Box 1 and Table 3 of the 2018 CDC guidance (CDC 2018) to arrive at the following group-to-tier mapping:

| Tier (High/Very High Severity) | subtier/rank within tier (for tier 1) | Group | Definition | CDC's Estimated Number in Group |
|---|---|---|---|---|
| 1 | 2 | Deployed and mission critical personnel for national security | Military forces and other mission critical personnel not limited to active duty military or U.S. government employees. Includes some diplomatic and intelligence service personnel, and public and private sector functions identified by federal agencies as unique and critical to national security | 850,000 |
| 1 | 7 | Public health personnel | Public health responders at federal, state and local levels | 300,000 |
| 1 | 1 for frontline, 7 for rest | Inpatient healthcare providers | Includes two-thirds of personnel at acute care hospitals who would be identified by their institution as critical to provision of inpatient health care services; primarily will include persons providing care with direct patient exposure but also will include persons essential to maintaining hospital infrastructure. | 3,200,000 |
| 1 | 4 for frontline 7 for rest | Outpatient and home health providers | Includes two-thirds of personnel identified by their organization at outpatient facilities, including but not limited to physicians' offices, primary care clinics, dialysis centers, urgent care centers, retail health clinics, and blood donation facilities; and skilled home health care personnel providing care with direct patient exposure. | 2,600,000 |
| 1 | 7 | Health care providers in long term care facilities (LTCFs) | Includes two-thirds of personnel at LTCFs identified by their organization as critical to the provision of care. | 1,600,000 |
| 1 | 4 | Pharmacists and pharmacy technicians | Includes pharmacists dispensing drugs at retail locations and pharmacy technicians who interact with the public and are essential for pharmacy operations (note that pharmacists in hospitals or outpatient centers may be targeted as part of those groups). | 725,000 |
| 1 | 3 for frontline EMS, 5 for frontline law enforcement and fire services, 7 for rest | Emergency services and public safety sector personnel (e.g. emergency medical service, law enforcement and fire services) | Includes groups supporting emergency response and public safety.  Emergency medical service personnel include those who are fire department based, hospital-based, or private; firefighters include professional and volunteers; law enforcement local police, sheriff officers, state troopers; and corrections officers include those in prisons and jails. | 2,000,000 |
| 1 | 7 | Manufacturers of pandemic vaccine and antiviral drugs | Includes critical personnel required for ongoing production of pandemic medical countermeasures to support a pandemic response. | 50,000 |
| 1 | 6 | Pregnant women | Women at any stage of pregnancy | 4,000,000 |
| 1 | 6 for age 6-11 months, 7 for age 1 or 2 | Infants and toddlers (6-35 months old) | Infants and toddlers in the specified age group | 11,000,000 |
| 2 | | Essential military support and sustainment personnel | Military and other essential personnel needed to support and sustain deployed forces | 650,000 |
| 2 | | Intelligence services | Critical personnel in the intelligence community serving at domestic and international posts that are not included in tier 1 | 150,000 |
| 2 | | National Guard personnel | National Guard Personnel not included in tier 1, but who are likely to be activated to maintain public order during a pandemic or to support pandemic response services or critical infrastructure | 500,000 |
| 2 | | Other domestic national security personnel | Includes other groups that are essential to national security such as guards at nuclear facilities and critical personnel providing border protection. | 150,000 |
| 2 | | Community support service personnel (emergency management and community and faith-based support and response organizations) | Personnel from community organizations who will provide essential support and have direct contact with persons and families affected during community pandemic outbreaks, and emergency management personnel who coordinate pandemic response and support activities | 600,000 |
| 2 | | Mortuary services personnel | Includes funeral directors | 50,000 |
| 2 | | Communications, information technology (IT), electricity, nuclear, oil & gas, and water sector personnel, financial clearing and settlement personnel | Personnel who are critical to support essential communications, information technology, utility, financial and other services provided by the defined sectors | 2,200,000 |
| 2 | | Critical government personnel – operational and regulatory functions | Federal, state, tribal, and tribal government employees and contractors who perform critical regulatory or operational functions required for essential operations of other critical infrastructure sectors | 425,000 |
| 2 | | Household contacts of infants under 6 months old. | Household contacts of infants under 6 months old | 4,500,000 |
| 2 | | Children 3 – 18 years old with a high risk medical condition | Children in the specified age group with a chronic medical condition that increases their risk of severe influenza disease, including heart and lung disease, metabolic disease, renal disease and neuromuscular diseases that may compromise respiratory function, as defined by ACIP recommendations for seasonal influenza vaccination. | 7,000,000 |
| 3 | | Remaining active duty military and essential support | Active duty personnel not included in higher priority groups and essential support personnel | 1,500,000 |
| 3 | | Other health care personnel | Includes groups that provide important health care services but are at less occupational risk, such as laboratory personnel | 350,000 |
| 3 | | Banking & finance, chemical, food & agriculture, pharmaceutical, postal & shipping, & transportation sector personnel. (Critical infrastructure with greater redundancy) | Personnel who are critical to support essential services provided by the defined sectors | 3,400,000 |
| 3 | | Other critical government personnel | Federal, state, tribal, and local government employees and contractors who perform important government functions included in agency continuity-of-operations plans | 400,000 |
| 3 | | Children 3 – 18 years old without a high risk medical condition | Children in the specified age group who do not have a chronic medical condition that increases their risk of severe influenza disease. | 62,000,000 |
| 4 | | Adults 19 – 64 years with a high risk condition | Adults in the specified age group with a chronic medical condition that increases their risk of severe influenza disease, including heart and lung diseases, metabolic diseases, renal disease, and neuromuscular diseases that may compromise respiratory function, as defined by ACIP recommendations for seasonal influenza vaccination | 38,000,000 |
| 4 | | Adults 65 years and older | Older adults in the specified age group | 41,000,000 |
| 5 | | Healthy adults, 19 – 64 years old | Adults in the specified age group not included above | 132,000,000 |

### A2.2. Data issues and imperfect workarounds

#### A2.2.1. Occupational specificity and supersetting

For the CDC's occupation-based groups (the national security, healthcare, and critical industries groups), the PUMS data does not have the necessary specificity to definitively identify someone as a member of the group. For example, one of the CDC's groups is manufacturers of pandemic vaccine and antiviral drugs. The PUMS data identifies those working in the pharmaceutical manufacturing industry but does not distinguish people making a COVID vaccine from people making non-COVID medicines.

Our work-around for the lack of required specificity needed to identify a specific group (such as manufacturers of pandemic vaccine and antiviral drugs) in the PUMS data is to construct a reasonable superset of the group (such as all those working in pharmaceutical manufacturing) and then randomly and independently assign each superset observation to the group with a fixed probability for the group. The fixed probability is chosen so that number of people in the US population represented by the observations assigned to the group approximately equal an externally provided estimate of number of people in the group.

The fixed probability is calculated as follows for assigning observations in a superset S to a group G that has an external size estimate of N:

$$P(observation \in G | observation \in S) = \frac{N}{\sum_{i \in S} PWGTP_i}$$

PWGTP, the variable representing the weight of each observation in the PUMS person data, is scaled so that $\sum_{i \in S} PWGTP_i$ serves as an estimator for the size of S in the US population. Subset sizes are all large relative to observation weights.

In the below tier-by-tier details, when we use this supersetting procedure to assign observations in a superset S to a group G that has an external size estimate of N, we will refer to it with the following: "Of the superset S, randomly assign N to group G."

We were not able to find a good external estimate of how many critical inpatient healthcare providers, outpatient healthcare providers, firefighters, EMS, firefighters, and law enforcement are frontline vs not frontline. The differences due to these values are confined to shifts between the subtiers of tier 1 and these values have no impact on cumulative demographics of vaccine recipients when there is enough vaccine for tier 1. As our reserve system does not kick in until tier 2, we have decided not to focus on defining frontline personnel and arbitrarily set the size of each frontline group to 0.5 the size of the critical group (e.g. our external estimate for the size of the group of frontline EMS is half the size of the group of critical EMS).

### A2.2.2. National security group size

The PUMS data seems to frequently greatly undercount the number of national security and military personnel in the population. Several times during our group identification exercise, we constructed what should have been a reasonable superset of a national security-based group only to have the total person weight of the PUMS superset be several times smaller than the CDC's estimate of the group size.

As we do not have access to the data that the CDC uses to come up with their numbers, we decided to go with the PUMS's numbers of national security personnel in these cases. More details can be found in the tier by tier criteria section of this document.

### A2.2.3. High-risk conditions and the BRFSS

The PUMS does not have data on the health conditions of the surveyed individuals. We therefore impute COVID-19 risk for each observation in the PUMS using the BRFSS data.

We label each observation in the BRFSS as high-risk for COVID-19 if at least one of the following is true:

- The individual has been "(Ever told) you had skin cancer"
- The individual has been "(Ever told) you had any other types of cancer"
- The individual has been "(Ever told) you have kidney disease" not including kidney stones, bladder infection or incontinence
- The individual has been "(Ever told) you have chronic obstructive pulmonary disease, C.O.P.D., emphysema or chronic bronchitis"
- The individual is obese, defined as having a BMI >= 30
- The individual has been "(Ever told) you had angina or coronary heart disease"
- The individual has been "(Ever told) you have diabetes," not including diabetes only while pregnant, pre-diabetes, or borderline diabetes

For each of the above statements, we consider each statement false if the variable takes any value other than the one listed, including values for "don't know/not sure," "not asked or missing," and "refused." The above variable definitions and question wordings are sourced from the 2018 BRFSS codebook (CDC 2018c).

The medical conditions chosen are an approximation of the CDC list of conditions (CDC 2020a) that make one of "increased risk of severe illness from COVID-19." As of 8/29/20, the conditions where people "are at increased risk," not including the conditions where people "might be at an increased risk," are as follows:

- Cancer
- Chronic kidney disease
- COPD (chronic obstructive pulmonary disease)
- Immunocompromised state (weakened immune system) from solid organ transplant
- Obesity (body mass index [BMI] of 30 or higher)
- Serious heart conditions, such as heart failure, coronary artery disease, or cardiomyopathies
- Sickle cell disease
- Type 2 diabetes mellitus

Our approximation of COVID-19 risk factors omits immunocompromised state from solid organ transplant and sickle cell disease due to lack of data in the BRFSS. Our risk factors also include all diabetes, not just type 2 diabetes, as the BRFSS does not distinguish between different types of diabetes.

Using the BRFSS data, labelled with whether or not there is high COVID-19 risk, we group the data by age bin, sex, whether or not the person is Hispanic, and race/ethnicity other than Hispanic. For each interaction of those variables, we find the proportion of the population that is high-risk. We omit any observations where one or more of these demographic variables are unknown or missing.

The age bins are the following:

- 18-24
- 25-29
- 30-34
- 35-39
- 40-44
- 45-49
- 50-54
- 55-59
- 60-64
- 65-69
- 70-74
- 75-79
- 80-84
- 85+

For non-Hispanic race, those of known race are categorized into the following:

- White only
- Black or African American only
- American Indian or Alaskan Native only
- Asian Only
- Native Hawaiian or other Pacific Islander only
- Other race only
- Multiracial

Hispanic status is coded in a separate variable from non-Hispanic race.

We considered adding income bins to the interaction of demographic variables, but the introduction of income resulted in lack of data for some variable interactions. We leave income out to avoid this and for simplicity.

We use to output the high-risk proportion of each demographic interaction in the BRFSS data. Then, we find the demographic probability of each PUMS person data observation being high risk based on the proportions calculated from the BRFSS. We then throw a weighted coin for each individual in the PUMS to label each individual as high risk or not. As the BRFSS does not have risk factor data for people under 18, we extrapolate the calculated risk probabilities of 18 to 24-year-olds of the same sex, race, and ethnicity for those under 18.

#### A2.2.4. Pregnant women

The PUMS data does not have data on whether or not each observation is of someone currently pregnant. It does denote whether or not a person has given birth in the past 12 months. We create a cohort of pregnant women in the data via the following steps:

1. Identify the entries in the PUMS data where:
    a. The person has given birth in the past 12 months AND
    b. We throw an independent weighted coin with heads probability equal to the length of a pregnancy as a fraction of a year and the coin comes up heads AND
    c. The observation is not included in any tier of a higher rank than tier 1 subtier 6.
2. Duplicate the entries identified in tier 1. Mark the duplicates but not the originals as pregnant. Mark all other entries as non-pregnant.
3. Subtract 1 from the age of the duplicates

The goal of this process is to minimize the differences when comparing the calculated demographics and size of each tier based on our synthesized dataset with the duplicated entries versus the calculated demographics and size of each tier based on a counterfactual PUMS dataset that included pregnancy data. We will refer to this counterfactual dataset in this section as "pregnancy-complete data" and the calculated values based on it as "pregnancy-complete data values".

Note that we could not have simply labelled those marked in step 1 as pregnant without creating a large and incorrect overlap between entries marked as pregnant and entries that are household contacts of a baby under 6 months. With the counterfactual pregnancy-complete data, there would have not been that overlap as most pregnant people intuitively would not also have a young infant in the house. This issue would cause many of those who would have been Tier 2 with pregnancy-complete data due to being household contacts of babies under 6 months would have instead been marked pregnant and counted with Tier 1, not Tier 2, potentially causing a noticeable difference from the pregnancy-complete data for the calculated demographic composition of Tier 2.

Our method of duplicating some entries and only marking the duplicates as pregnant has the effect that our calculated demographics and sizes of tiers strictly higher than tier 1 subtier 6 are identical to those tiers' pregnancy-complete values. For tiers strictly lower than tier 1 subtier 6, the difference between the dataset generated with our method and the pregnancy-complete data is that those that do not qualify for any tier strictly higher than tier 1 subtier 6 AND would have been known to be pregnant in the pregnancy-complete data are included in the lower tier's demographic summary with our method and excluded in the complete data values. As long as pregnant women compose a small minority of each lower tier, our method minimizes deviations from the pregnancy-complete data values for all higher and lower tiers.

The demographics and size of tier 1 subtier 6 itself can be reasonably considered similar to the pregnancy-complete data values. The synthetic cohort of pregnant women are of a size and racial composition comparable to the true number of pregnant women by nature of our large-scaled randomization as long as there is little year-to-year variation in pregnancy counts and demographics and little in- and out-flow of pregnant women and new mothers in the US.
The length of pregnancy used in our calculation was 268 days, referencing Jukic et al. (2013). The length of a year used in our calculation was 365 days, referencing simplicity and common sense.

We note that our count of pregnant women using only the ACS (the number of people who gave birth in the last year, multiplied by the length of a pregnancy as a fraction of a year) is only about 3 million rather than the CDC's provided 4 million. We do not know the reason for this discrepancy.

### A2.3. Comment on tier and group sizes

Except for the national security groups affected by the counting issue, our numbers for the sizes of each occupation-based group are tied to the CDC estimates of group size via the supersetting procedure.

We do not align the numbers for high-risk demographics with the CDC's numbers, as they're either fully identifiable or imputable with ACS data (e.g. age-based and pregnancy) or differ for COVID-19 rather than influenza such that CDC numbers do not apply (e.g. high-risk status). Our numbers for the sizes of the tiers differ from the CDC estimates of tier size for 2 reasons:

1. The sizes of our demographic-based groups are not tied to CDC estimates and often differ.
2. We remove those belonging to a higher tier from the count for each tier. As we understand, the CDC does not.

### A2.4. Tier by tier criteria details for labelling of PUMS data

#### A2.4.1. National security personnel

*A2.4.1.1. Tier 1*

- Deployed and mission critical personnel for national security (tier 1 subtier 2)
    - CDC size estimate: 850k
    - Of the superset of active duty military personnel randomly assign 200k to group as "deployed personnel." 200k as an estimate of the number of deployed personnel comes from Gibbons-Neff and Schmitt (2019).
    - Of the superset of observations that have not been assigned as deployed and have a NAICS code of 928P .ADM-National Security And International Affairs, randomly assign 650k (850k – 200k) as mission critical personnel

*A2.4.1.2. Tier 2*

- Critical intelligence services and other domestic national security personnel
    - This is the 2$^{nd}$ and 4$^{th}$ groups in Tier 2 combined
    - CDC combined size estimate: 300k
    - Superset: people in the NAICS industry code of 928P .ADM-National Security And International Affairs who have not already been included in Deployed and mission critical personnel for national security (tier 1 subtier 2)
    - Randomly assign 300k of this superset to this combined group
- National guard personnel not included in tier 1
    - CDC size estimate: 500k

- We attempted to use the superset of those with a NAICS code of 928110P6 .MIL-U.S. Armed Forces, Branch Not Specified or 928110P7 .MIL-Military Reserves Or National Guard who have not already been assigned to tier 1 subtier 2. However, this set only represents 72k people in the US population, significantly less than the CDC's estimate of 500k. As we could not increase our number without including people from other branches of the military and non-military individuals, we decided to assign everyone in our attempted superset to this group instead of going through our supersetting procedure.
- Essential military support and sustainment personnel
    - CDC size estimate: 650k
    - Superset: those with a military NAICS code (928110P1 .MIL-U.S. Army, 928110P2 .MIL-U.S. Air Force, 928110P3 .MIL-U.S. Navy, 928110P4 .MIL-U.S. Marines, 928110P5 .MIL-U.S. Coast Guard, 928110P6 .MIL-U.S. Armed Forces, Branch Not Specified, 928110P7 .MIL-Military Reserves Or National Guard) who have not been assigned to any of the above national security groups
    - Randomly assign 650k of this superset to this group

*A2.4.1.3. Tier 3*

- remaining active duty military and essential support
    - CDC size estimate: 1.5 million
    - We attempted to use the superset of those not assigned to a higher-ranking national security group who are active duty military personnel (MIL == 1) OR who have a military NAICS code (928110P1 .MIL-U.S. Army, 928110P2 .MIL-U.S. Air Force, 928110P3 .MIL-U.S. Navy, 928110P4 .MIL-U.S. Marines, 928110P5 .MIL-U.S. Coast Guard, 928110P6 .MIL-U.S. Armed Forces, Branch Not Specified, 928110P7 .MIL-Military Reserves Or National Guard). However, this set only represents 254k people in the US population, significantly less than the CDC's estimate of 1.5 million. As we could not increase our number without including non-military individuals, we decided to assign everyone in our attempted superset to this group instead of going through our supersetting procedure.

### A2.4.2. Healthcare personnel

*A2.4.2.1. Tier 1*

- critical inpatient healthcare providers (tier 1 subtier 1 if frontline, 1 subtier 7 if not)
    - CDC size estimate: 3.2 million
    - Superset: those with a NAICS code of 622M MED-General Medical And Surgical Hospitals, And Specialty (Except Psychiatric And Substance Abuse) Hospitals
    - Randomly assign 3.2 million of the superset to the group
    - Of those assigned in this group, randomly assign half to tier 1 subtier 1 as frontline inpatient HCWs and half to tier 1 subtier as non-frontline inpatient HCWs.
- critical outpatient healthcare providers (tier 1 subtier 4 if frontline, tier 1 subtier 7 if not)
    - CDC size estimate:2.6 million

- Superset: those with a NAICS code of NAICS of 6211 MED-Offices Of Physicians OR 6214 MED-Outpatient Care Centers OR 6216 MED-Home Health Care Services
    - Randomly assign 2.6 million of the superset to the group.
    - Of those assigned in this group, randomly assign half to tier 1 subtier 4 as frontline outpatient HCWs and half to tier 1 subtier 7 as non-frontline outpatient HCWs.
- critical long-term care facility (LTCF) personnel (tier 1 subtier 7)
    - CDC size estimate:1.6 million
    - Superset: those with a NAICS code of 6231 MED-Nursing Care Facilities (Skilled Nursing Facilities) OR 623M MED-Residential Care Facilities, Except Skilled Nursing Facilities
    - Randomly assign 1.6 million of the superset to the group
- public health personnel (tier 1 subtier 7)
    - CDC size estimate:300k
    - Superset: those in a NAICS industry in the MED category and have not already assigned to any a previous healthcare group
        - MED category: "6211", "6212", "62131", "62132", "6213ZM", "6214", "6216", "621M", "622M", "6222", "6231", "623M"
        - There is no industry code for public health, so we are randomly labelling members of the medical industry as public health personnel.
    - Randomly assign 300k of the superset to the group

*A2.4.2.2. Tier 3*

- Other healthcare personnel
    - CDC size estimate:350k
    - Superset: those in a NAICS industry in the MED category and have not already assigned to any a previous healthcare group
        - MED category: "6211", "6212", "62131", "62132", "6213ZM", "6214", "6216", "621M", "622M", "6222", "6231", "623M"
    - Randomly assign 350k of the superset to the group

### A2.4.3. Employees of critical industries

*A2.4.3.1. Tier 1*

- critical pharmacists and pharmacy technicians (tier 1 subtier 4)
    - CDC size estimate:725k
    - Superset: those with a SOC occupation code of 291051 MED-Pharmacists or 292052 MED-Pharmacy Technicians
    - Randomly assign 725k of the superset to this group
- critical EMS, law enforcement, and fire services (tier 1 subtier 3 for frontline EMS; tier 1 subtier 5 for frontline law enforcement and frontline fire services; tier 1 subtier 7 for non-frontline critical EMS, law enforcement, and fire services)
    - CDC size estimate: 2 million
    - Superset: those with a SOC code of one of the following:

- 292042 MED-Emergency Medical Technicians
        - 292043 MED-Paramedics OR
        - 533011 TRN-Ambulance Drivers And Attendants, Except Emergency Medical Technicians
        - 331011 PRT-First-Line Supervisors Of Correctional Officers
        - 331012 PRT-First-Line Supervisors Of Police And Detectives
        - 331021 PRT-First-Line Supervisors Of Fire Fighting And Prevention Workers
        - 333050 PRT-Police Officers
        - 333012 PRT-Correctional Officers And Jailers
        - 332011 PRT-Firefighters
    - Randomly assign 2 million of the superset to this group
    - For those that have been assigned to this group and are EMS (SOC code of "292042", "292043", "533011"), randomly assign each to tier 1 subtier 3 as frontline or tier 1 subtier 7 as non-frontline with probability ½.
    - For those that have been assigned to this group and are law enforcement or firefighters (SOC code of "331011", "331012", "331021", "333012", "333050", "332011"), randomly assign each to tier 1 subtier 5 as frontline or tier 1 subtier 7 as non-frontline with probability ½.
- critical antivirals/vaccine manufacturers (tier 1_7)
    - CDC size estimate: 50k
    - Superset: those with NAICS code of 3254 MFG-Pharmaceuticals And Medicines
    - Randomly assign 50k of the superset to this group

*A2.4.3.2. Tier 2*

- community support service personnel (emergency management and community and faith-based support and response organizations)
    - CDC size estimate: 600k
    - Superset: those with a NAICS code of one of the following:
        - 6241 SCA-Individual And Family Services
        - 6242 SCA-Community Food And Housing, And Emergency Services
        - 6243 SCA-Vocational Rehabilitation Services
        - 6244 SCA-Child Day Care Services
    - Randomly assign 600k of the superset to this group
- mortuary services personnel
    - CDC size estimate: 50k
    - Superset: those with a SOC code of one of 394031 .PRS-Morticians, Undertakers, And Funeral Arrangers OR 3940XX .PRS-Embalmers, Crematory Operators, And Funeral .Attendants
    - Randomly assign 50k of the superset to this group
- Communications, information technology (IT), electricity, nuclear, oil & gas, and water sector personnel, financial clearing and settlement personnel
    - CDC size estimate: 2.2 million
    - Superset: those with a NAICS code of one of the following:
        - 211 .EXT-Oil And Gas Extraction
        - 2211P .UTL-Electric Power Generation, Transmission And .Distribution,
        - 2212P .UTL-Natural Gas Distribution,

- 22132 .UTL-Sewage Treatment Facilities,
        - 2213M .UTL-Water, Steam, Air Conditioning, And Irrigation Systems,
        - 221MP .UTL-Electric And Gas, And Other Combinations,
        - 22S .UTL-Not Specified Utilities,
        - 517311 .INF-Wired Telecommunications Carriers,
        - 517Z .INF-Telecommunications, Except Wired Telecommunications .Carriers,
        - 522M .FIN-Nondepository Credit And Related Activities
    - Randomly assign 2.2 million of the superset to the group
- critical government personnel - operational and regulatory functions
    - CDC size estimate: 425k
    - Superset: those with a NAICS code of one of the following:
        - 92113 .ADM-Public Finance Activities,
        - 92119 .ADM-Other General Government And Support,
        - 9211MP .ADM-Executive Offices And Legislative Bodies,
        - 923 .ADM-Administration Of Human Resource Programs,
        - 92M1 .ADM-Administration Of Environmental Quality And Housing Programs,
        - 92M2 .ADM-Administration Of Economic Programs And Space Research,
        - 92MP .ADM-Justice, Public Order, And Safety Activities
    - Randomly assign 425k of the superset to the group

*A2.4.3.3. Tier 3*

- Banking & finance, chemical, food & agriculture, pharmaceutical, postal & shipping, & transportation sector personnel
    - CDC size estimate: 3.4 million
    - Superset: those who were not assigned as critical vaccine/drug manufacturers (tier 1 critical industries) or financial clearing personnel (tier 2 critical industries) AND have one of the following NAICS codes:
        - 5221M .FIN-Savings Institutions, Including Credit Unions,
        - 522M .FIN-Nondepository Credit And Related Activities,
        - 5241 .FIN-Insurance Carriers,
        - 5242 .FIN-Agencies, Brokerages, And Other Insurance Related .Activities, 52M1 .FIN-Banking And Related Activities,
        - 52M2 .FIN-Securities, Commodities, Funds, Trusts, And Other .Financial Investments,,
        - 111 .AGR-Crop Production,
        - 112 .AGR-Animal Production And Aquaculture,
        - 115 .AGR-Support Activities For Agriculture And Forestry,
        - 3254 .MFG-Pharmaceuticals And Medicines,
        - 481 .TRN-Air Transportation,
        - 482 .TRN-Rail Transportation,
        - 483 .TRN-Water Transportation,
        - 484 .TRN-Truck Transportation,
        - 4853 .TRN-Taxi And Limousine Service,
        - 485M .TRN-Bus Service And Urban Transit,

- 486 .TRN-Pipeline Transportation,
        - 488 .TRN-Services Incidental To Transportation,
        - 491 .TRN-Postal Service, 492 .TRN-Couriers And Messengers,
        - 493 .TRN-Warehousing And Storage,
        - 3252 .MFG-Resin, Synthetic Rubber, And Fibers And Filaments,
        - 3253 .MFG-Agricultural Chemicals,
        - 3254 .MFG-Pharmaceuticals And Medicines,
        - 3255 .MFG-Paint, Coating, And Adhesives,
        - 3256 .MFG-Soap, Cleaning Compound, And Cosmetics,
        - 325M .MFG-Industrial And Miscellaneous Chemicals
    - Randomly assign 3.4 million of the superset to the group
- Other critical government personnel
    - CDC size estimate: 400k
    - Superset: those who were not assigned as part of the tier 2 critical government personnel group AND who have a NAICS code of one of the following:
        - 92113 .ADM-Public Finance Activities,
        - 92119 .ADM-Other General Government And Support,
        - 9211MP .ADM-Executive Offices And Legislative Bodies,
        - 923 .ADM-Administration Of Human Resource Programs,
        - 92M1 .ADM-Administration Of Environmental Quality And Housing Programs,
        - 92M2 .ADM-Administration Of Economic Programs And Space Research,
        - 92MP .ADM-Justice, Public Order, And Safety Activities
    - Randomly assign 400k of the superset to the group

#### A2.4.4. High-risk demographics

Note that our labelling procedures for these groups do not involve the CDC estimates of group size.

##### *A2.4.4.1. Tier 1*

- Pregnant women (tier 1 subtier 6)
    - Assign observations labelled as pregnant via the procedure described in the "Pregnant women" section of this document to this group.
- Infants age 6-11 months (tier 1 subtier 6)
    - For every individual with an age of 0, independently and randomly assign to either age 0-5 months or 6-11 months.
    - Assign those labelled 6-11 months to this group.
- Toddlers age 1-2 years (tier 1 subtier 7)
    - Assigned individuals with an age of 1 or 2 to this group.

##### *A2.4.4.2. Tier 2*

- household contacts of infants under 6 months
    - Identify and assign to this group those sharing a household ID number (SERIALNO) with an individual labelled as age 0-5 months via our Infants age 6-11 months identification procedure above.

- children age 3-18 with a high-risk medical condition
    - Assign to this group individuals age 3 to 18, inclusive, who have been labelled as having a high-risk condition via the procedure described in the "High-risk conditions and the BRFSS" section of this document.

*A2.4.4.3. Tier 3*

- children 3-18 without a high-risk medical condition
    - Assign to this group individuals age 3 to 18, inclusive, who have NOT been labelled as having a high-risk condition via the procedure described in the "High-risk conditions and the BRFSS" section of this document.

*A2.4.4.4. Tier 4*

- Adults 19-64 with a high-risk condition
    - Assign to this group individuals age 19-64, inclusive, who have been labelled as having a high-risk condition via the procedure described in the "High-risk conditions and the BRFSS" section of this document.
- Adults 65 and older
    - Assign to this group individuals with age >= 65.

## A3. Calculation of ADI Values

### A3.1. Overview

We calculate raw numerical ADI values using the 1990 components and weighing formula from Singh (2003). Table 1 of Singh (2003), copied below, defines each component in the "Census Variable" column and provides each component's weights in the "Factor Score Coefficient" column.

**TABLE 1—Factor Loadings and Factor Score Coefficients for the Census Variables Comprising the Area Deprivation Index: United States, 1970 and 1990**[21]

| Census Variable | Factor Loading: Tract Index, 1990 | Factor Loading: Zip Code Index, 1990 | Factor Loading: County Index, 1990 | Factor Score Coefficient: Tract Index, 1990 | Factor Loading: County Index, 1970 |
|---|---|---|---|---|---|
| Population aged ≥ 25 y with < 9 y of education, %[a] | 0.7498 | 0.7383 | 0.7885 | 0.0849 | 0.8340 |
| Population aged ≥ 25 y with at least a high school diploma, % | −0.8562 | −0.8089 | −0.8231 | −0.0970 | −0.8788 |
| Employed persons aged ≥ 16 y in white-collar occupations, % | −0.7721 | −0.7118 | −0.6890 | −0.0874 | −0.6075 |
| Median family income, $ | −0.8629 | −0.8690 | −0.9218 | −0.0977 | −0.8694 |
| Income disparity[b] | 0.8262 | 0.7054 | 0.8827 | 0.0936 | 0.7559 |
| Median home value, $ | −0.6074 | −0.6764 | −0.6740 | −0.0688 | −0.6703 |
| Median gross rent, $ | −0.6896 | −0.7081 | −0.7876 | −0.0781 | −0.7872 |
| Median monthly mortgage, $ | −0.6795 | −0.7362 | −0.7812 | −0.0770 | ... |
| Owner-occupied housing units, % (home ownership rate) | −0.5431 | −0.4688 | −0.4408 | −0.0615 | ... |
| Civilian labor force population aged ≥16 y unemployed, % (unemployment rate) | 0.7117 | 0.5231 | 0.5679 | 0.0806 | 0.2195 |
| Families below poverty level, % | 0.8623 | 0.7996 | 0.8796 | 0.0977 | 0.9480 |
| Population below 150% of the poverty threshold, % | 0.9157 | 0.8781 | 0.9266 | 0.1037 | 0.9503 |
| Single-parent households with children aged < 18 y, % | 0.6346 | 0.3487 | 0.3329 | 0.0719 | 0.5520 |
| Households without a motor vehicle, % | 0.6126 | 0.4335 | 0.4549 | 0.0694 | 0.7540 |
| Households without a telephone, % | 0.7748 | 0.6837 | 0.7830 | 0.0877 | 0.8745 |
| Occupied housing units without complete plumbing, % (log) | 0.4505 | 0.4863 | 0.6392 | 0.0510 | 0.8921 |
| Households with more than 1 person per room, % (crowding) | 0.4910 | 0.3963 | 0.4018 | 0.0556 | 0.6854 |
| Proportion of total variance explained by factor | 0.5195 | 0.4432 | 0.5140 | ... | 0.5990 |
| Cronbach α (reliability coefficient) | 0.9484 | 0.9311 | 0.9473 | ... | 0.9573 |

*Note.* Values were derived from a principal-components analysis of census ecological data for 59 525 census tracts, 29 320 zip codes, and 3097 counties.
[a]For the 1970 index, percentage of population with less than 5 years of education was used.
[b]Income disparity in 1990 was defined as the log of 100 × ratio of number of households with < $10 000 income to number of households with ≥$50 000 income. Income disparity in 1970 was defined as the log of 100 × ratio of number of households with < $3000 income to number of households with ≥$15 000 income.

The raw numerical ADI score of each family is calculated as the weighted sum of its value for each component using the weights above. For a family F, the raw numerical ADI is as follows:

$$ADI_F^{raw} = \sum_{component} value_{component,F} * weight_{component}$$

We then assign the raw numerical ADI of each family to each member of the family. We estimate the raw numerical ADI distribution of the entire U.S., treating each observation i in the PUMS person data as $PWGTP_i$ people in the US population. We then divide this national distribution into 10 national deciles and label them from 1 to 10, where 10 represents the highest ADI. We define an observation in the PUMS as high-ADI as belonging to a national ADI decile, or simply "having an ADI", of >= 8.

### A3.2. Missing values and omitted observations

Most of the PUMS variables we use are not counted for the ~8 million living in group quarters, which include prisons, nursing facilities, and college/university housing (US Census Bureau 2017). We do not compute a raw numerical ADI for those people and those people are excluded from the deciles. People in group quarters are automatically excluded from the high-ADI for reserve purposes. High-ADI is defined as being in the top 3 deciles of people who do not live in group quarters. Thus, the proportion of people designated as high-ADI is slightly less than 30%.

We note that the actual ADI computed at the census block level also omits computing an ADI for census blocks with more than 33% of the population living in group quarters (University of Wisconsin School of Medicine and Public Health 2015).

For those living in households, any missing components will be considered 0. Every household has at least 1 missing component, as property value and mortgage are missing for all renters and gross rent is missing for all who own the houses they live in.

For averages computed at the family level, we consider the household of those living in a household with a non-missing family income variable to be a family. This includes households headed by same-sex couples. For those not living with family due to living by themselves or with non-family roommates, we consider each person to be their own family.

### A3.3. Component by component details

1. Population aged ≥25 y with < 9 y of education, %
   a. Calculate the family proportions of those 25 and over in the family whose highest educational attainment is under grade 9. Families with no member over 25 have this value missing.
2. Population aged ≥25 y with at least a high school diploma, %
   a. Calculate the family proportions of those 25 and over in the family whose have graduated high school, have a GED, or completed at least some college. Families with no member over 25 have this value missing.
3. Employed persons aged ≥16 y in white-collar occupations, %
   a. Calculate the family proportion of those over 16 and employed whose NAICS industry code is in the MGR, BUS, FIN, CMM, ENG, SCI, CMS, LGL, EDU, ENT, MED, SAL, or OFF categories. Families with no employed member over 16 have this value missing.
4. Median family income, $
   a. Use the family income adjusted to constant dollars. For those not living with family, use individual income adjusted to constant dollars.
5. Income disparity
   a. Skipped and set to 0 for all, due to data constraints
6. Median home value, $
   a. Use household property value adjusted to constant dollars. Non-homeowners have this value missing.
7. Median gross rent, $
   a. Use monthly gross rent adjusted with to constant dollars. Non-renters have this value missing.
8. Median monthly mortgage, $
   a. Use the first mortgage payment adjusted with to constant dollars. Non-homeowners and those who did not mortgage have this value missing.
9. Owner-occupied housing units, %
   a. Home ownership is 100% if the house is owned by its resident and 0% if otherwise
10. Civilian labor force population aged ≥16 y unemployed, %

a. Calculate the family proportion of those over 16 in the civilian labor force who are unemployed. This component is missing for families whose members over 16 are all in the armed forces or not in the labor force.
11. Families below poverty level, %
    a. For households containing families, 100% if family-income-to-poverty percentage < 100%. For individuals not living with family, compare unadjusted individual income to poverty thresholds of the sample's year (U.S. Census Bureau 2020b).
12. Population below 150% of the poverty threshold, %
    a. For households containing families, 100% if family-income-to-poverty percentage < 150. For individuals not living with family, compare unadjusted individual income to 150% of poverty thresholds of the sample's year (U.S. Census Bureau 2020b).
13. Single-parent households with children aged < 18 y, %
    a. 100% if the household is a family not headed by a same sex married couple and with a man and no wife or a woman and no husband with related children under 18. 0 otherwise.
14. Households without a motor vehicle, %
    a. 100% if at least 1 vehicle available
15. Households without a telephone, %
    a. 100% if no telephone service AND no cell data plan for smartphone or other mobile device
16. Occupied housing units without complete plumbing, %
    a. 100% if there are complete plumbing facilities
17. Households with more than 1 person per room, %
    a. 100% if the number of people in the household is greater than the number of rooms.

## A4. Notes on Graphing

### A4.1. Death rates (actual and age-adjusted) by race/ethnicity

The 8/5/20 release of APM Research Lab's Color of Coronavirus data only provides actual death rates and age-adjusted death rates for the racial categories of Indigenous, Asian, Black, Latino, White, and Pacific Islander, omitting actual death rates and age-adjusted death rates for those who are multiracial, some other race, or unknown race.
Therefore, to approximate the actual proportion of deceased Covid-19 victims who are Black or Indigenous, we use the following formula:

$$\widehat{deathshare}_{BI} = \frac{population_B * deathrate_B + population_I * deathrate_I}{\sum_{race\,=\,I,A,B,L,W,PI} population_{race} * deathrate_{race}}$$

This equals the true value of Black or Indigenous share of Covid-19 actual deaths with the following sufficient assumptions:

1. APM Research Lab's values for each race's share of population and actual death rate are accurate given their assumptions, and data deficiencies (APM Research Lab 2020).

2. Of the deceased victims counted by APM Research Labs as multiracial, some other race, or unknown race, the proportion who identified as Black or Indigenous is equal to $\widehat{deathshare}_{BI}$

The procedure and assumptions for calculating age-adjusted death share for Black and Indigenous people and the two death rates for Black, Indigenous, and Latino people are analogous.

### A4.2. State by state distribution

State maps were created as Excel filled maps with a "Diverging (3-color)" color scheme with the "Midpoint" (yellow) value set at 1.0. The "Maximum" (green) and "Minimum" (red) values are displayed at the top and bottom of the legend, respectively.

The estimated vaccines received per state under the various reserve systems was calculated in our main procedures above**.** The population per state was computed from the ACS data with their weights**.** The "fair share" index relative to population is then the ratio of the proportion of the total vaccines received by a state to the proportion of population living in a state.

To calculate analogous fair share indices relative to case and death incidence, first data on total confirmed cases and deaths in each state was extracted on August 14 (CDC 2020b). Then the ratio of cases in each state relative to total cases in the country, and deaths in each state relative to total death sin the country, were calculated. The fair share index is then the ratio of the proportion of total vaccines a state received to the proportion of total national cases each state recorded. This was done analogously for deaths.

All three of these fair share indices were calculated at different total supply levels of vaccines (e.g. 20 million, 50 million), and for different reserve systems (0% reserve, 20% reserve, 40% reserve).

## A5. Table of Program Files

| File name | Description |
|---|---|
| personal_adi_calculation.R | Assigns ADI to PUMS data. |
| brfss.R | Calculates proportion of each demographic group that has a high-risk condition, using the BRFSS. |
| pums_tier_labelling.R | Assigns all qualifying CDC tiers to PUMS data. |
| tier_demographic_summary.R | Summarizes demographic characteristics of population by highest tier and ADI |
| tier_demographic_graphing.R | Implements 2018 CDC priority system or ADI reserve system and outputs recipient characteristics. |
| racereserve_demographic_summary.R | Summarizes demographic characteristics of population by highest tier and Black/Indigenous status |
| racereserve_demographic_graphing.R | Implements Black/Indigenous reserve system and compares results to ADI reserve system. |
| New State Analysis.xlsx | Creates maps and graphs for state-by-state allocations. |
| cumulative counts of high ADI recipients over n_vaccinesMarginal.xlsx | Calculates the high-ADI proportion of the marginal vaccine recipient. |
| ColorOfCoronavirus_Data_File_ ThroughAug4_2020_APMResearchLab.xlsx | Calculates proportion of COVID deaths and age-adjusted deaths attributed to each race. |

**References for the Appendix**